\documentclass[12pt]{article} 
\usepackage{graphicx,amsmath, amssymb, verbatim}
\usepackage[all]{xy}
\newtheorem{definition}{Definition}
\newtheorem{example}{Example}
\newtheorem{conjecture}{Conjecture}
\newtheorem{corollary}{Corollary}
\newtheorem{remark}{Remark}

\newcommand {\C } {\mathbb{C}} 

\newcommand {\p } {\mathbb{P}}

\newcommand {\Z} {\mathbb{Z}} 
\DeclareMathOperator{\Res}{\mathrm{Res}}

\newcommand{\ba}{\begin{eqnarray}}
\newcommand{\ea}{\end{eqnarray}}
\newcommand{\no}{\nonumber}

\begin{document} 

\title{\bf{Prepotentials for local mirror symmetry via Calabi-Yau fourfolds
}}

\author{Brian Forbes,\; Masao Jinzenji \\ \\ \it Division of 
Mathematics, Graduate School of Science \\ \it Hokkaido University \\ 
\it 
Kita-ku, Sapporo, 060-0810, Japan \\ \it brian@math.sci.hokudai.ac.jp\\
\it jin@math.sci.hokudai.ac.jp }

\maketitle 

\begin {abstract} 
In this paper, we first derive an intrinsic definition of classical triple 
intersection numbers of $K_{S}$, where $S$ is a complex toric  
surface, and use this to compute the extended Picard-Fuchs system of $K_{S}$ of 
\cite{FJ}, without making use of the instanton expansion. We then extend this formalism to local fourfolds $K_{X}$, where $X$ is a complex 3-fold. As a result, we are able to fix the prepotential of local Calabi-Yau threefolds $K_S$ up to polynomial terms of degree 2. We then outline methods of extending the procedure to non canonical bundle cases.  

\end {abstract} 

\section{Introduction.}

The basic utility of mirror symmetry is its power in the computation of Gromov-Witten invariants. In terms of classical, compact mirror symmetry, these invariants are computed from the coefficients of a generating function, known as the prepotential. What one does in practice is solve for the period integrals of the mirror manifold, and then identify the prepotential and mirror map as certain linear combinations of ratios of these period integrals.

In the context of local mirror symmetry, in which one considers mirror symmetry for noncompact Calabi-Yau manifolds, such an approach has not appeared to date. In fact, current technology does not provide us with a means of defining the prepotential in these cases. This problem emerges because the local mirror manifold does not have `enough' period integrals to determine the prepotential. While we can often turn to localization formulas to determine Gromov-Witten invariants, this is generally more cumbersome than the corresponding $B$ model calculation. The above problem becomes manifest in the case of the mirror computation 
of $K_{F_{2}}$, using the double log solution of the usual Picard-Fuchs system. 
As is well known, the Hirzebruch surface $F_{2}$ has one Calabi-Yau 
direction in the two dimensional K\"ahler cone. Therefore, local Gromov-Witten 
invariants associated with curves that have 
positive degree only in this direction cannot be computed from the double log 
solution, as was suggested in \cite{CKYZ}. In \cite{FJ}, we proposed the idea 
of an extended Picard-Fuchs system for local mirror symmetry, obtained by
modifying the usual Picard-Fuchs system of local mirror symmetry. The extended
Picard-Fuchs system has a larger solution space than the usual one, and moreover it has a triple log solution. Therefore, we can compute the full prepotential of a local 
Calabi-Yau threefold. In particular, in the case of $K_{F_{2}}$, the triple log 
solution includes the information of local Gromov-Witten 
invariants that cannot be detected by the double log solution! However, a basic problem with the constructions of \cite{FJ} is that the instanton expansion of the prepotential was used to derive the extended system. In the case of $K_{S}$, where $S$ is a
compact toric surface, this instanton data fixes the 
triple intersection numbers, which are crucial in the construction of the extended 
Picard-Fuchs system. For $X$ a local Calabi-Yau 3-fold with $\dim H_{4}
(X,\Z)=0$, we had to make direct use of the instanton part of the $A$-model 
prepotential to derive an extended Picard-Fuchs system.        

The aim of this paper is to overcome these weak points in the construction of the 
extended Picard-Fuchs system. In the case of $K_{S}$, we derive a natural 
definition of the classical triple intersection numbers of $K_{S}$, by 
generalizing the definition of local Gromov-Witten invariants given in 
\cite{CKYZ}. This definition matches the results in \cite{FJ} and explains 
the moduli parameter of the classical triple intersection numbers found 
in \cite{FJ}. Therefore, we can construct an extended Picard-Fuchs system 
of $K_{S}$ without using the instanton part of the prepotential of $K_{S}$.  

On the other hand, we may also take advantage of our formula for intersection theory in order to provide an alternate derivation of the prepotential of local mirror symmetry. In the event that $X=K_S$, the procedure goes as follows. First, we construct a compact 3-fold by taking the projective closure of $X$ : ${\bar X}
=\p(\mathcal O_S\oplus K_S)$. We then consider 
the local Calabi-Yau 4-fold $K_{\bar X}$ and construct an extended Picard-Fuchs system of $K_{\bar X}$ by generalizing the results for $K_{S}$. With this extended Picard-Fuchs system, we can compute the 4-point Yukawa couplings of $K_{\bar X}$. Finally, we can see the instanton part of the three point functions of $X$ by taking the large fiber limit of the fourpoint functions.
This provides a simple algorithm by which we can extract the exact form of the prepotential for $K_S$ (up to polynomial terms of degree 2).
  
A problem with this approach appears if the dimension of the compactification fiber gets too large. In particular, we run into this difficulty for any noncompact threefold $X$ such that $\dim H_2(X,\Z)\ge 2$, $\dim H_4(X,\Z)=0$. For such examples, we provide a method by which one may reduce the problem to a $K_S$ case. Then, in the appropriate limits, we are again able to give a definition for the prepotential.

The organization of this paper is as follows.
In Section 2, we first propose a conjecture on a geometrical 
interpretation of the 3 point functions of local mirror symmetry for $K_{S}$, which is a 
straightforward generalization of the definition given in \cite{CKYZ}, and use this to 
derive a formula for classical triple intersection numbers for $K_{S}$. Then 
we compute explicitly these numbers for the examples used in \cite{FJ} from the 
localization formula.  In Section 3, we extend the results of the previous 
section to the local fourfold $K_{\bar X}$ and construct the extended Picard 
Fuchs system of $K_{\bar X}$ in the case of $X=K_{S}$. Next, we clarify the relation between the large fiber limit of the 4 point functions of $K_{\bar X}$ and the 3 point functions of $X$. We also justify the process of the computation of the B-model 4-point functions by using the extended Picard-Fuchs system.    
Section 4 gives the toric construction of the projective closure $\bar X$
when $X$ is a vector bundle. 
   Section 5 contains applications of the fourfold construction to $K_S$ and the total space of $\p^1$ with normal bundle $\mathcal O(-1)\oplus \mathcal O(-1)$ or $\mathcal O\oplus \mathcal O(-2)$. Section 6 details methods of dealing with more exotic cases. Finally, the extended Picard-Fuchs system (in the sense of \cite{FJ}) for the trivalent curve is given in the appendix.

\bigskip

$\bf{Acknowledgments.}$ 
 
The research of B.F. was 
funded by a COE grant of Hokkaido University. The research of M.J. is 
partially 
supported by JSPS grant No. 16740216.

\bigskip 

\section{Fractional intersection theory on $K_S$.}

We begin by discussing fractional intersection theory for noncompact Calabi-Yau threefolds \cite{JN}\cite{FJ}. The discussion here will be necessary for fixing the form of the extended Picard-Fuchs system, and will eventually allow us to fix the overall scaling factor of the prepotential exactly. 

\subsection{A conjecture on Yukawa couplings.}
In our previous paper \cite{FJ}, we computed the Yukawa couplings (3-point functions)
of a local Calabi-Yau 3-fold $K_{S}$ ($S$: toric 2-fold) by using the extended 
Picard-Fuchs system. In this subsection, we first write down a conjecture 
on the geometrical interpretation of these Yukawa couplings:
\begin{conjecture}
The Yukawa couplings computed in \cite{FJ} are the three point functions \\
$\langle {\cal O}_{J_{a}}(z_{1}){\cal O}_{J_{b}}(z_{2}){\cal O}_{J_{c}}(z_{3})
\rangle_{0}$ of the topological sigma model on $K_{S}$ without coupling to 
topological gravity:
\begin{eqnarray}
\langle {\cal O}_{J_{a}}(z_{1}){\cal O}_{J_{b}}(z_{2}){\cal O}_{J_{c}}(z_{3})
\rangle_{0}&=&\sum_{\vec{d}}q^{\vec{d}}
\langle {\cal O}_{J_{a}}(z_{1}){\cal O}_{J_{b}}(z_{2}){\cal O}_{J_{c}}(z_{3})
\rangle_{0,\vec{d}}\nonumber\\
&:=&\sum_{\vec{d}}q^{\vec{d}}
\int_{[\overline{M}_{0}(S,\vec{d})]_{vir.}}
\frac{c_{top}(R^{1}\pi_{*}ev^{*}K_{S})}
{c_{top}(R^{0}\pi_{*}ev^{*}K_{S})}ev_{1}^{*}(J_{a})ev_{2}^{*}(J_{b})ev_{3}^{*}(J_{c}).
\label{st}
\end{eqnarray}
\end{conjecture}
Here, we have to explain the notation used in (\ref{st}). We denote the generators of $H^{1,1}(S,\Z)$ by $J_{a}$. $\overline{M}_{0}(S,\vec{d})$ is the compactified moduli space of holomorphic maps of degree $\vec{d}\in H_{2}(S,\Z)$ from $\mathbb{P}^{1}$ to $S$. The notation $[\overline{M}_{0}(S,\vec{d})]_{vir.}$ means that we always insert the top Chern class of the obstruction bundle in the same way as in the usual theory of Gromov-Witten invariants. 
We note that this moduli space does not correspond to the topological sigma model coupled to topological gravity. Therefore, 
we don't take the equivalence class of $SL(2, \C)$, the 
automorphism group of $\mathbb{P}^1$, and we also don't consider the degrees of freedom 
from moving marked points. Instead, we introduce three fixed marked points 
$z_{1},z_{2},z_{3}\in \mathbb{P}^{1}$ and define the evaluation maps $ev_{i}:\overline{M}_{0}(S,\vec{d}) \rightarrow S$ by $\varphi(z_{i})\in S,\;\;(\varphi\in \overline{M}_{0}(S,\vec{d}))$. We also define the map $ev: \mathbb{P}^{1}
\times \overline{M}_{0}(S,\vec{d})\rightarrow S$ by $ev(z,\varphi)=\varphi(z)$
and the map $\pi:\mathbb{P}^{1}
\times \overline{M}_{0}(S,\vec{d})\rightarrow \overline{M}_{0}(S,\vec{d})$
as the projection map onto the second factor.   

In the case of a local 3-fold $K_{S}$, we have a birational map 
between $\overline{M}_{0}(S,\vec{d})$ and the usual moduli space of 
stable maps $\overline{M}_{0,3}(S,\vec{d})$, because $SL(2,\C)$ is 
isomorphic to the position of the three distinguished marked points in 
$\mathbb{P}^{1}$. Therefore, this definition coincides with the usual definition of 3-point local Gromov-Witten invariants of $K_{S}$, at least in the case 
when $\vec{d}\neq {0}$ \cite{CKYZ}. As is well known, the extension of this 
conjecture to higher dimensional local Calabi-Yau manifolds is slightly 
different from the usual theory of local Gromov-Witten invariants. 

In our previous paper \cite{FJ}, a crucial point of the construction of the extended 
Picard-Fuchs system of $K_{S}$ is the determination of the classical part of the 
Yukawa couplings. Therefore, we carefully look at the $\vec{d}=0$ part of the 
above conjecture. In this case, $\varphi\in\overline{M}_{0}(S,0)$ is just 
the constant map from $\mathbb{P}^{1}$ to $S$, and it is obvious that
$\overline{M}_{0}(S,0)=S$. Hence $ev_{i}$ turns out to be the identity map of 
$S$. The map $ev$ becomes a projection map of the second factor of 
$\mathbb{P}^{1}\times \overline{M}_{0}(S,0)$, and $R^{1}\pi_{*}ev^{*}K_{S}$
and $R^{0}\pi_{*}ev^{*}K_{S}$ turn out to be $0$ and $K_{S}$ respectively. 
With these considerations, the classical triple intersection number
$\langle C_{a}, C_{b}, C_{c}\rangle:=\langle {\cal O}_{J_{a}}(z_{1}){\cal O}_{J_{b}}(z_{2}){\cal O}_{J_{c}}(z_{3})
\rangle_{0,0}$ ($C_{a}\in H_{4}(S,\Z)$ is the Poincare dual of $J_{a}$) 
is given by the formula: 
\begin{corollary}
\begin{equation}
\langle C_{a}, C_{b}, C_{c}\rangle=
\int_{S}\frac{J_{a}J_{b}J_{c}}{c_{1}(K_{S})}.
\label{cor}
\end{equation}
\end{corollary}
At first glance, this formula seems to be ill-defined, because division 
by $c_{1}(K_{S})$ is not defined in $H^{*}(S,\C)$. Yet $K_{S}$ is written 
in terms of a linear combination of $J_{a}$'s, and we can therefore expect 
the following constraints between classical triple intersection numbers
by the formal reduction $c_{1}(K_{S})/c_{1}(K_{S})=1$:
\begin{corollary}
\begin{equation}
\langle C_{a}, C_{b}, PD(c_{1}(K_{S}))\rangle=\int_{S}J_{a}J_{b},
\label{const}
\end{equation}
\end{corollary}
where we denote  the Poincare dual of $c_{1}(K_{S})$ by $PD(c_{1}(K_{S}))$ .
In the case of $K_{\mathbb{P}^{2}}$, $H^{1,1}(\mathbb{P}^{2},\Z)$ is 
generated by the hyperplane class $H$, and the above corollary gives us
\begin{equation} 
\langle H, H, H\rangle=\int_{\mathbb{P}^2}\frac{H^{3}}{-3H}=-\frac{1}{3},
\end{equation}
which coincides with the result in \cite{CKYZ}. 
In the next subsection, we try to compute the r. h. s. of (\ref{cor}) with the 
aid of the localization formula, and we also show that the constraint 
(\ref{const}) holds in the results obtained in our previous paper \cite{FJ}. 

As another application of the above conjecture, we compute the three point 
function of $K_{F_{2}}$ that corresponds to the non-rigid curve in the fiber 
direction. This computation has already been mentioned in \cite{CKYZ}, but 
it is important in our fourfold construction that will be introduced in the 
next section. Let us first introduce the toric construction of $F_{2}$. 
$F_{2}$ is obtained from dividing 
$\C^{4}\setminus(((0,0)\times\C^{2})\cup(\C^{2}\times(0,0)))$ by the two 
$\C^{*}$ actions,
\begin{equation}
(x_{1},x_{2},x_{3},x_{4})
\sim(x_{1},x_{2},\mu x_{3},\mu x_{4})
\sim(\lambda x_{1},\lambda x_{2},x_{3},\lambda^{-2}x_{4}).
\label{f2}
\end{equation}
The classical cohomology of $F_{2}$ is generated by the two K\"ahler forms 
$J_{u}$ and $J_{v}$ that correspond to the $\mu$ and $\lambda$ 
actions respectively. These K\"ahler forms satisfy the following relations:
\begin{equation}
J_{u}^{2}=2J_{u}J_{v},\;\;\;J_{v}^{2}=0.
\label{f2c}
\end{equation}
Then we consider two holomorphic maps with degrees $(d_{u},d_{v})=
(1,0)$ and $(d_{u},d_{v})=(0,1)$, as follows:
\begin{eqnarray}
\varphi_{1}(s,t)&=&(a,b,c_{1}s+c_{2}t,d_{1}s+d_{2}t),\nonumber\\
\varphi_{2}(s,t)&=&(a_{1}s+a_{2}t,b_{1}s+b_{2}t,c,0).
\label{hol}
\end{eqnarray}
Note that the fourth entry of $\varphi_{2}$ should be $0$ because of the 
$\lambda^{-2}$ action. By considering the two $\C^{*}$ actions, 
we can see that moduli space of $\varphi_{1}$ and $\varphi_{2}$ 
can be compactified into $\mathbb{P}^{1}\times\mathbb{P}^{3}$ 
and $\mathbb{P}^{3}$ respectively. Therefore, the image curve 
of $\varphi_{1}$ is not rigid in $F_{2}$, but the image curve 
of $\varphi_{2}$ is rigid in $F_{2}$. Next, we extend this construction 
to $K_{F_{2}}$. $K_{F_{2}}$ is constructed by adding a fifth variable 
$x_{5}$ and extending the two $\C^{*}$ actions as follows:
\begin{equation}
(x_{1},x_{2},x_{3},x_{4},x_{5})
\sim(x_{1},x_{2},\mu x_{3},\mu x_{4},\mu^{-2}x_{5})
\sim(\lambda x_{1},\lambda x_{2},x_{3},\lambda^{-2}x_{4},x_{5}).
\label{f2}
\end{equation}     
Then the two holomorphic maps in (\ref{hol}) can be extended to
\begin{eqnarray}
\tilde{\varphi}_{1}(s,t)&=&(a,b,c_{1}s+c_{2}t,d_{1}s+d_{2}t,0),\nonumber\\
\tilde{\varphi}_{2}(s,t)&=&(a_{1}s+a_{2}t,b_{1}s+b_{2}t,c,0,e).
\label{hol2}
\end{eqnarray}
We note here that the fifth entry of $\tilde{\varphi}_{1}(s,t)$ 
should be $0$ by the $\mu^{-2}$ action. Therefore, we can conclude that the
image curve of $\tilde{\varphi}_{1}(s,t)$ is rigid along the non-compact
fiber direction, as in the usual situation in local mirror 
symmetry. But $\tilde{\varphi}_{2}(s,t)$ has one additional moduli 
parameter $e$, which corresponds to a non-compact fiber direction.
This situation is exceptional, and so we compute the three point function 
$\langle {\cal O}_{J_{v}}(z_{1}){\cal O}_{J_{v}}(z_{2}){\cal O}_{J_{v}}(z_{3})
\rangle_{(0,1)}$ for the degree $(0,1)$ map 
$\tilde{\varphi}_{2}(s,t)$ by the following. 
If we look back at our conjecture, the appearance of 
the additional moduli parameter $e$ results in the non-trivially of
$c_{top}(R^{0}\pi_{*}ev^{*}(K_{F_{2}}))$, and this turns out to be 
$-2J_{u}$ in this case. On the other hand, 
$R^{1}\pi_{*}ev^{*}(K_{F_{2}})$ is trivial, so what remains to be computed is
\begin{equation}
\langle {\cal O}_{J_{v}}(z_{1}){\cal O}_{J_{v}}(z_{2}){\cal O}_{J_{v}}(z_{3})
\rangle_{(0,1)}
=\int_{[\overline{M}(F_{2},(0,1))]_{vir.}}\frac{1}{-2J_{u}}
ev_{1}^{*}(J_{v})ev_{2}^{*}(J_{v})ev_{3}^{*}(J_{v}).
\end{equation} 
This formula seems exotic, but luckily, we have a nontrivial 
virtual fundamental class in this case. Since the
normal bundle $N$ of the image curve in $F_{2}$ is generated by $x_{4}$,
it is isomorphic to ${\cal O}_{F_{2}}(-2J_{v}+J_{u})$. Therefore, 
$\varphi_{2}^{*}N$ is identified with ${\cal O}_{\mathbb{P}^{1}}(-2)
\otimes {\cal O}_{F_{2}}(J_{u})$ and we have 
$c_{top}(R^{1}\pi_{*}ev^{*}(N))=J_{u}$. Hence, we have obtained the following 
equality:
\begin{eqnarray}
\langle {\cal O}_{J_{v}}(z_{1}){\cal O}_{J_{v}}(z_{2}){\cal O}_{J_{v}}(z_{3})
\rangle_{(0,1)}
&=&\int_{[\overline{M}(F_{2},(0,1))]_{vir.}}\frac{1}{-2J_{u}}
ev_{1}^{*}(J_{v})ev_{2}^{*}(J_{v})ev_{3}^{*}(J_{v})\nonumber\\
&=&\int_{\overline{M}(F_{2},(0,1))}\frac{J_{u}}{-2J_{u}}
ev_{1}^{*}(J_{v})ev_{2}^{*}(J_{v})ev_{3}^{*}(J_{v})\nonumber\\
&=&-\frac{1}{2}\int_{\mathbb{P}^{3}}
H^{3}\nonumber\\
&=&-\frac{1}{2},
\end{eqnarray} 
where we used the results that follow from the previous compactification:
\begin{equation}
\overline{M}(F_{2},(0,1))=\mathbb{P}^{3},\;\;\;ev_{i}(J_{v})=H
,\;\;(\mbox{$H$ is the hyperplane class of $\mathbb{P}^{3}$}).
\end{equation} 
As was suggested in \cite{CKYZ}, this fractional Gromov-Witten invariant 
cannot be seen from the usual recipe of local mirror symmetry, which relies 
on one double log solution. But we can detect this invariant by the extended 
Picard-Fuchs system of $K_{F_{2}}$ constructed in \cite{FJ}, since 
this system has a triple log solution. This fact is one of the non-trivial 
advantages of the extended Picard-Fuchs system.
 
\subsection{Review of the fixed point formula.}

In this part, we will review the application of the Atiyah-Bott fixed point formula to torically described surfaces $S$, where the number of independent curve classes $C \in H_2(S,\Z)$ is allowed to be arbitrary.  The Hirzebruch surface $F_2$ will be used as an example throughout this discussion.

So, let $S$ be a smooth toric complex twofold, defined by vertices $\{\nu_1,\dots,\nu_n\}\subset \Z^m$ and a choice of basis $\{l^1,\dots,\l^{n-2}\} \subset \Z^n$ of relations for the $\nu_i$. That is, if $l^j=(l^j_1,\dots,l^j_n)$, then $\sum_{i=1}^n l^j_i\nu_i=0$ for all $j$. We note, in particular, that smooth toric varieties are simplicial. Recall (see e.g. \cite{CK}) that to each $v_i$ there is an associated divisor $D_i \in H_2(S,\Z)$, and similarly, to each $l^j$ we may associated a curve class $C_j \in H_2(S,\Z)$. Moreover, the intersection matrix between these divisors and curves is determined by 
\begin{equation}
D_i \cdot C_j = l^j_i.
\end{equation}

For a more tangible view of $S$ and its curves and divisors, we can use the homogeneous coordinate ring representation \cite{C}. This gives an isomorphism
\begin{equation}
S \cong \frac{\C^n-Z}{(\C^*)^{n-2}}
\end{equation}
where $Z$ is the Stanley-Reisner ideal, and the action of the $j$th factor of the quotient appears as
\begin{equation}
\C^*:(x_1,\dots,x_n)\longrightarrow(\alpha^{l^j_1}x_1,\dots,\alpha^{l^j_n}x_n).
\end{equation}
$\alpha$ is the generator of $\C^*$. If $(x_1,\dots,x_n)$ are coordinates on $\C^n$, we can then simply describe the divisors of $S$ by $D_i = S \cap \{x_i=0\}.$ 

In the case of $F_2$, we have vertices
\begin{eqnarray}
\nu_1=(1,0), \nu_2=(0,1), \nu_3=(-1,2),\nu_4=(0,-1)	
\end{eqnarray}
The a basis of relations for these is provided by
\begin{equation}
I_{ij}=
\begin{pmatrix}
 l^1 \\ l^2   
\end{pmatrix}=
\begin{pmatrix}
1 & 1 & 0 & 0   \\ -2 & 0 & 1 & 1  
\end{pmatrix}
\end{equation}
and, as mentioned above, $I_{ij}=D_i \cdot C_j.$ We also have $Z=\{x_1x_2=0\} \cup \{x_3x_4=0\}$.

To apply the localization formula, it is convenient to first compute the equivariant cohomology ring of $S$. To construct this, begin with the ordinary cohomology ring
\begin{eqnarray}
H^*(S,\C)=\frac{\C[K_1,\dots,K_n]}{\big(P,Z(K_i)\big)}.
\end{eqnarray}
The $K_i$ are the Poincare duals of the divisors $D_i$, and $P$ is the ideal of linear relations for the $K_i$. $Z(K_i)$ is the Stanley Reisner ideal, where  $x$ has been replaced by $K$. For the curve classes $C_j$ defined by the basis vectors of relations among the vertices $\nu_i$, we introduce K\"ahler classes $J_i \in H^{1,1}(S,\C)$ such that
\begin{eqnarray}
\int_{C_j}J_i=\delta_{ij}.	
\end{eqnarray}
The cohomology classes  $K_j$ and $J_i$ are related in a very simple way; namely
\begin{eqnarray}
K_i=\sum_k l^k_iJ_k.	
\end{eqnarray}
We are now in position to write down the equivariant cohomology ring of $S$ with respect to the group action $T$ on $S$ inherited from $\C^n$; it is
\begin{eqnarray}
H^*_T(S,\C)=\frac{\C[J_1,\dots,J_{n-2},\lambda_1,\dots,\lambda_n]}{Z(\sum_k l^k_iJ_k-\lambda_i)}.	
\end{eqnarray}

Let $\{p_1,\dots,p_r\}$ be the fixed points of the action $T$ on $S$. Recall that in this situation, if $i_j:p_j\hookrightarrow S$ is the inclusion map and $N_j=N_{p_j/S}$, then the fixed point formula reads
\begin{eqnarray}
\int_{S_T}\nu= \sum_{j=1}^r\frac{i^*_j(\nu)}{e_T(N_j)}.	
\end{eqnarray}
Above, $\nu\in H_T^*(S)\otimes \C[\lambda_1,\dots,\lambda_n]$, $e_T(N_j)$ is the equivariant Euler class of $N_j$, and if $ET\rightarrow BT$ is the classifying bundle of $T$, then $S_T=S\times_T ET.$ 

To apply this formula, it is useful to have an algorithm for the computation of $e_T(N_j)$. This can be readily done, as follows. First write $Z(\sum_k l^k_iJ_k-\lambda_i)=\{R_1(J,\lambda)\dots R_{\alpha}(J,\lambda)\},$ where we are using the shorthand $J=(J_1\dots J_{n-2}), \lambda=(\lambda_1\dots \lambda_n)$. In our setting, each factor $R_i(J,\lambda)$ breaks down as a product of linear factors $P^i_k$:
\begin{eqnarray}
	R_i(J,\lambda)=\prod_{j=1}^{n_i} P^i_j(J,\lambda).
\end{eqnarray}
where $\prod_{i=1}^{\alpha} n_i=r$.
Then solving the relations $R_1(J,\lambda)=\dots=R_{\alpha}(J,\lambda)=0$ for $J$ in terms of $\lambda$, we find $r$ solutions. Without loss of generality, we use the first solution for the purpose of this explanation, which can be described by 
\begin{eqnarray}
	P^1_1(J,\lambda)=\dots=P^{\alpha}_1(J,\lambda)=0.
\end{eqnarray}
Let $J(\lambda)$ denote the solution to this equation. 
Then the equivariant Euler class of the normal bundle is given by the formula
\begin{eqnarray}
e_T(N_1)=\prod_{i=1}^{\alpha} 	\prod_{j=2}^{n_i} P^i_j(J(\lambda),\lambda)
\end{eqnarray}
We obtain similar formulas for each of the other $r-1$ solutions.

We now apply this to $F_2$. The intersection matrix $I_{ij}$ tells us that the ordinary cohomology ring of $F_2$ can be written
\begin{eqnarray}
H^*(F_2,\C)=\frac{\C[K_1,\dots,K_4]}{\langle K_3-K_4,K_1+K_3+K_4-K_2,K_1K_2,K_3K_4 \rangle}.	
\end{eqnarray}
Thus the equivariant cohomology ring is given as
\begin{eqnarray}
H^*_T(F_2,\C)=\frac{\C[J_1,J_2,\lambda_1,\dots,\lambda_4]}{\langle(J_1-2J_2-\lambda_1)(J_1-\lambda_2),(J_2-\lambda_3)(J_2-\lambda_4) \rangle}	.
\end{eqnarray}
One of the solutions of the relations of the Stanley Reisner ideal is $J_1=2\lambda_4+\lambda_1,J_2=\lambda_4$. Substituting this into the remaining nonzero terms, we find
\begin{eqnarray}
	e_T(N)=(\lambda_2-2\lambda_4-\lambda_1)(\lambda_4-\lambda_3)
\end{eqnarray}
for the equivariant Euler class of the normal bundle at this fixed point. There are exactly 4 such fixed points using this construction, as expected. 

As a test of these calculations, we can compute the intersection numbers between the 2-cycles on $F_2$ via the fixed point theorem. Then we find e.g.
\begin{eqnarray*}
C_1\cdot C_2=\int_{F_2}J_1\wedge J_2 \
= \ \frac{\lambda_2\lambda_4}{(\lambda_4-\lambda_3)(\lambda_2-2\lambda_4-\lambda_1)}	+\frac{\lambda_2\lambda_3}{(\lambda_3-\lambda_4)(\lambda_2-2\lambda_3-\lambda_1)}	+\\ \frac{(2\lambda_4+\lambda_1)\lambda_4}{(\lambda_4-\lambda_3)(\lambda_1+2\lambda_4-\lambda_2)}	+\frac{(2\lambda_3+\lambda_1)\lambda_3}{(\lambda_3-\lambda_4)(\lambda_1+2\lambda_3-\lambda_2)}=1,	
\end{eqnarray*}
the correct intersection number.

We then give the general definition, based on our conjecture 
in the previous section:
\begin{definition}
\label{intnumbs}
 Let $S$ be a toric surface with torus action $T$, and let $\{p_1,\dots,p_r\}$ be the isolated fixed points of $T$ on $S$. Let $C_a,C_b,C_c\in H_2(S,\Z)$, and denote the canonical bundle of $S$ by $K_S$. Then the \normalfont triple intersection numbers of $S$ \it are defined by the formula
\begin{eqnarray}
	\left\langle C_a, C_b,C_c \right\rangle=\sum_{j=1}^r\frac{i^*_j(J_a)i^*_j(J_b)i^*_j(J_c)}{e_T(N_j)}i^*_j(e_T^{-1}(K_S)).
\end{eqnarray}
Here $i_j:p_j\hookrightarrow S$ is the inclusion, $N_j$ is the normal bundle of $p_j$ in $S$ and $e_T(E)$ denotes the equivariant Euler class of the bundle $E$. Also the $J_i$ satisfy $\int_{C_i}J_j=\delta_{ij}$. 
\end{definition}
This definition is a precise version of the heuristic formula for triple intersection numbers derived earlier: 
\begin{eqnarray}
\label{heuristic}
	\left\langle C_a, C_b,C_c \right\rangle=\int_S \frac{J_aJ_bJ_c}{c_1(K_S)}. 
\end{eqnarray}
While Definition 1 is mathematically rigorous, in practice it can be cumbersome to write out the sometimes quite complicated formulas of the torus weights. As such, we can make use of the formula (\ref{heuristic}) to make a heuristic calculation of the intersection numbers. This is in fact the strategy we will employ when computing intersection numbers for the del Pezzo surface below. 

\subsection{Examples.}

\begin{example} \normalfont

Let's first use the definition on a rather simple case, namely $F_0=\p^1\times \p^1$. From section 2.1, we have that the equivariant cohomology ring of $F_0$ with respect to the standard $T$ action is 
\begin{eqnarray}
H^*_T(F_0,\C)=\frac{\C[J_1,J_2,\lambda_1,\dots,\lambda_4]}{\langle(J_1-\lambda_1)(J_1-\lambda_2),(J_2-\lambda_3)(J_2-\lambda_4) \rangle}.	
\end{eqnarray}
Note that there are four fixed points $p_1,\dots,p_4$ corresponding to the four corners of the square $\p^1\times \p^1$. Then we can use the above expression for the equivariant cohomology to find the inverse images of the two cohomology classes $J_1,J_2$, as well as of the canonical bundle. We write out one of the expressions we get by using the above definition:
\begin{eqnarray}
\left\langle C_1, C_1, C_2  \right\rangle = \frac{\lambda_1^2\lambda_3}{(\lambda_3-\lambda_4)(\lambda_1-\lambda_2)(-2\lambda_1-2\lambda_3)}
+\frac{\lambda_1^2\lambda_4}{(\lambda_4-\lambda_3)(\lambda_1-\lambda_2)(-2\lambda_1-2\lambda_4)} \\ \no	
+\frac{\lambda_2^2\lambda_3}{(\lambda_3-\lambda_4)(-\lambda_1+\lambda_2)(-2\lambda_2-2\lambda_3)}	+\frac{\lambda_2^2\lambda_4}{(\lambda_4-\lambda_3)(-\lambda_1+\lambda_2)(-2\lambda_2-2\lambda_4)}.		
\end{eqnarray}
There are, naturally, three others for the other triple intersection numbers. Then all we need to do is set $\lambda_1=\lambda_3$ and we immediately have that 
\begin{eqnarray}
\left\langle C_1, C_1, C_1  \right\rangle =\frac{x}{4}, \ \left\langle C_1, C_1, C_2  \right\rangle=-\frac{x}{4}, \ \left\langle C_1, C_2, C_2  \right\rangle =\frac{x-2}{4}, \ \left\langle C_2, C_2, C_2  \right\rangle =\frac{2-x}{4} 	
\end{eqnarray}
where $x$ is an expression involving the torus weights, which we interpret here as a moduli parameter on the intersection numbers.
These are exactly the four triple intersection numbers from \cite{JN}\cite{FJ}
\footnote{The extended Picard-Fuchs system of $K_{F_{0}}$ indeed has one 
moduli parameter which agrees with the above results, but it was not 
mentioned in \cite{FJ}.}.

\end{example}

\begin{example}\normalfont
Next, consider $F_2$. Here we will find that we must make a nontrivial choice of torus weights in order to reproduce the expected triple intersection numbers \cite{FJ}. The origin of this complication lies in the fact that the canonical bundle over $F_2$ does not involve a cohomology class from the base curve. In \cite{FJ}, this ambiguity turned up as a moduli parameter for the intersection numbers. 

As above, we find three of the four triple intersection numbers on $F_2$. Note that here, the values computed are independent of the choice of torus weights, in contrast to the $F_0$ case. 
\begin{eqnarray}
\left\langle C_1, C_1, C_1  \right\rangle =-1, \ \left\langle C_1, C_1, C_2  \right\rangle=-\frac{1}{2}, \ \left\langle C_1, C_2, C_2  \right\rangle =0. 	
\end{eqnarray}
Again, these agree with \cite{FJ}. However, for the remaining intersection number we obtain
\begin{eqnarray}
\left\langle C_2, C_2, C_2  \right\rangle	=\frac{1}{2}\frac{\lambda_1\lambda_3^2+\lambda_1\lambda_3\lambda_4+\lambda_1\lambda_4^2+2\lambda_3\lambda_4^2+2\lambda_3^2\lambda_4}{(2\lambda_3+\lambda_1)(2\lambda_4+\lambda_1)\lambda_2}
\end{eqnarray}
At first,
this result seems to mean that there exists one moduli parameter corresponding to $\left\langle C_2, C_2, C_2  \right\rangle$. However, in \cite{FJ}, we have found that
this number should be set to zero from considering the behavior of the triple log solution of the extended Picard-Fuchs system \cite{FJ}. We think that this phenomena is deeply connected with the exceptional behavior of the curve in $K_{F_{2}}$ that is nonrigid in the noncompact direction. 
\end{example}
\begin{example}\normalfont
We can also carry out the calculation for $F_1$. 
The equivariant cohomology ring is in this case
\begin{eqnarray}
	H_T^*(F_1,\C)=\frac{\C[J_1,J_2,\lambda_1,\dots,\lambda_4]}{\langle (J_1-\kappa_1)(J_1-J_2-\kappa_2),(J_2-\kappa_3)(J_2-\kappa_4)\rangle}
\end{eqnarray}
From \cite{FJ}, it was found that there is in fact a moduli parameter in the triple intersection numbers for this case which leaves the instanton expansion invariant. Using the localization calculation, this problem shows up as an indeterminacy of the intersection numbers. However, what we find is that by fixing one of the four intersection numbers, the other three are determined automatically. We fix $\left\langle C_1, C_1, C_1  \right\rangle =x$ by choosing \begin{equation}
\lambda_3=\frac{-\lambda_1(3\lambda_1+2\lambda_4+12x\lambda_1+6x\lambda_4)}{(1+3x)(2\lambda_1+\lambda_4)}, \ x \ne -\frac{1}{3}.
\end{equation}
 Then this choice gives the remaining three intersection numbers 
\begin{eqnarray}
 \left\langle C_1, C_1, C_2  \right\rangle=-1-2x, \ \left\langle C_1, C_2, C_2  \right\rangle =1+4x, \ \left\langle C_2, C_2, C_2  \right\rangle =-2-8x.	
\end{eqnarray}
These are again as expected, including the moduli parameter \cite{FJ}.
\end{example}
\begin{example} \normalfont
Finally, we compute triple 
intersection numbers for the del Pezzo surface $dP_2$. In this case, the 
fixed-point computation
 is rather complicated, and we therefore  present an alternative (simplified) way of 
determining the classical triple intersection numbers. First, we restate 
the notation of the previous paper \cite{FJ} for the classical 
cohomology ring of $dP_{2}$.
It is generated by three K\"ahler forms $J_1,\;J_2,\;J_3$ 
and obeys the 5 relations:
\begin{eqnarray}
&&p_1=(J_1-J_2)(J_1-J_3),\;p_2=J_2(J_2+J_3-J_1),\;
p_3=J_3(J_2+J_3-J_1),\no\\
&&p_4=J_2(J_1-J_3),\;p_5=J_3(J_1-J_2).
\end{eqnarray}
As in the previous examples, the $J_i$ are chosen such that if $C_1,C_2,C_3$ is a basis of $H_2(dP_2,\Z)$, then 
\begin{eqnarray}
	\int_{C_i}J_j=\delta_{ij}.
\end{eqnarray}
With this notation, $e(N):=c_{1}(K_{dP_{2}})$ is given by $-(J_1+J_2+J_3)$, and the triple intersection numbers are given by the formula:
\begin{equation}
\langle C_a, C_b, C_c\rangle=\int_{dP_2}\frac{J_aJ_bJ_c}{c_1(K_{dP_2})}= -\int_{dP_{2}}\frac{J_a
J_bJ_c}{J_1+J_2+J_3}.  
\end{equation}
Of course, the above expression is formal, but we can read off from this 
equation the relations between triple intersection numbers:
\begin{equation}
\langle C_1, C_b, C_c\rangle+\langle C_2, C_b, C_c\rangle+\langle C_3, C_b, C_c\rangle=
-\int_{dP_{2}}J_bJ_c.
\label{lr}
\end{equation}
Notice that the r.h.s is just the well-defined classical intersection number of 
$dP_{2}$. Since the classical triple intersection numbers are symmetric 
in $a,\;b,\;c$, we have 10 independent numbers. But (\ref{lr}) imposes 
6 independent relations between these numbers. As a result, we obtain 4 
moduli parameters in the classical triple intersection numbers, which  
agree with the 4 moduli parameters found in the previous paper \cite{FJ}.     
\begin{eqnarray}
&&\langle C_1, C_1, C_1\rangle=-1+3x+3z+y+w,\;\;
\langle C_2, C_2, C_2\rangle=-y,\;\;
\langle C_3, C_3, C_3\rangle=-w,\nonumber\\
&&\langle C_1, C_1, C_2\rangle=-z-2x-y,\;\;
\langle C_1, C_1, C_3\rangle=-x-2z-w,\nonumber\\
&&\langle C_1, C_2, C_2\rangle=x+y,\;\;
\langle C_1, C_3, C_3\rangle=z+w,\nonumber\\
&&\langle C_2, C_2, C_3\rangle=-x,\;\;
\langle C_2, C_3, C_3\rangle=-z,\;\;
\langle C_1, C_2, C_3\rangle=x+z-1.
\label{cldP}
\end{eqnarray}
If we set $z=x,\;w=y$, these results reduce to the triple intersection 
numbers used in \cite{FJ}.
\end{example}
\begin{remark}
In the case of $K_{F_{0}}, K_{F_{1}}$ and $K_{dP_{2}}$, the extended 
Picard-Fuchs system found from the instanton part of the prepotential 
has the same number of moduli parameters which cannot be fixed by the 
constraints (\ref{const}), but the extended Picard-Fuchs system 
of $K_{F_{2}}$ has no moduli. As we have mentioned, this fact seems to
be related to the existence of a $(0,1)$ curve in $K_{F_{2}}$ which is 
not rigid along the fiber direction.  
\end{remark}

\section{The computational strategy via local fourfold.}

In this section, we give a schematic presentation of the procedure we will be using to determine the prepotential. Let $X$ be a noncompact Calabi-Yau threefold, and let $\mathcal F$ denote the prepotential for $X$, which we want to define by using mirror symmetry. For all cases considered in this paper, $X$ is either $K_S$, the canonical bundle over a complex surface $S$, 
or $\dim H_4(X,\Z)=0.$

First, let us suppose that $X=K_S$. Then we compute $\mathcal F$ in the following steps:
\bigskip
\newline
\indent 1) Take the canonical bundle over the projective closure of $K_S$, $\hat X=\mathcal O(K)\rightarrow \p(\mathcal O_S \oplus K_S)$. 
\bigskip
\newline
\indent 2) Let $\hat Y$ be the mirror of $\hat X$, and compute the fourpoint functions of $\hat Y$ using the extended 
\newline \indent Picard-Fuchs system.
\bigskip
\newline
\indent 3) Using the mirror map, convert the fourpoint functions of $\hat Y$ into fourpoint functions of $\hat X$.
\bigskip
\newline
\indent 4) Recover $\mathcal F$ from the fourpoint functions of $\hat X$ in the large fiber limit. 
\newline
\bigskip

At this point, we include a brief discussion as to why we expect to be able to derive $\mathcal F$ from the above steps. After all, the resulting manifolds $\hat Y$ and $\hat X$ are still noncompact, and we may therefore find ourselves in the same situation as in the original noncompact $K_S$ case. The main point, however, is that while $\hat Y$ is a noncompact fourfold, it contains the compact threefold $\p(\mathcal O_S \oplus K_S)$ as a submanifold. Then, as in the considerations of \cite{CKYZ}, they found that they were able to derive Picard-Fuchs equations for spaces like $K_S$, but that the resulting differential systems corresponded to the underlying compact twofold $S$. Similarly, with our procedure we expect to be able to derive accurately all information corresponding to the underlying compact threefold $\p(\mathcal O_S \oplus K_S)$ from $\hat Y$.  

We now give a more detailed explanation of the above steps. 
Let $\{J_1,\dots,J_m\}$ be a basis of $H^{1,1}(X,\C)$. Take the projective closure $\bar X=\p(\mathcal O_S \oplus K_S)$ of $X$, and let $J_F$ be a K\"ahler class of $\bar X$ such that $\{J_1,\dots,J_m,J_F\}$ is a basis of $H^{1,1}(\bar X,\C)$.  We consider the canonical bundle over $\bar X$: $\hat X= K_{\bar X}$. Since $\hat X$ is a Calabi-Yau fourfold, the idea is then to compute $\mathcal F$ as a limit of the fourpoint functions $C_{ijkl}$ of $\hat X$. At this stage, 
we clarify the geometrical meaning of the fourpoint functions 
$C_{ijkl}$ of the local 4-fold $\hat X=K_{\bar X}$. This can be done by a 
direct generalization of our conjecture given in the previous 
section. 
\begin{conjecture}
The fourpoint function $C_{ijkl}$ is the fourpoint function 
$\langle {\cal O}_{J_{i}}(z_{1}){\cal O}_{J_{j}}(z_{2})
{\cal O}_{J_{k}}(z_{3}){\cal O}_{J_{l}}(z_{4})
\rangle_{0}$ of the topological sigma model on $K_{\bar X}$ without coupling to 
topological gravity:
\begin{eqnarray}
C_{ijkl}&=&\sum_{\vec{d}}q^{(\vec{d},d_{F})}
\langle {\cal O}_{J_{i}}(z_{1}){\cal O}_{J_{j}}(z_{2}){\cal O}_{J_{k}}(z_{3})
{\cal O}_{J_{l}}(z_{4})
\rangle_{0,(\vec{d},d_{F})}\nonumber\\
&:=&\sum_{\vec{d}}q^{(\vec{d},d_{F})}
\int_{[\overline{M}_{0}({\bar X},(\vec{d},d_{F}))]_{vir.}}
\frac{c_{top}(R^{1}\pi_{*}ev^{*}K_{\bar X})}
{c_{top}(R^{0}\pi_{*}ev^{*}K_{\bar X})}ev_{1}^{*}(J_{i})ev_{2}^{*}(J_{j})ev_{3}^{*}(J_{k})ev_{4}^{*}(J_{l}),
\label{st2}
\end{eqnarray}
where the notation is the same as that in Conjecture 1.
\end{conjecture}

We compute the $C_{ijkl}$ by using mirror symmetry. Let $\hat Y$ be the Hori-Vafa mirror to $\hat X$, and take $\{z_1,\dots, z_m, z_F\}$ to be complex structure coordinates for $\hat Y$, where $z_F$ is mirror to the complexified K\"ahler coordinate $t_F$ satisfying $\Re (t_F)=J_F$. We first determine the fourpoint functions $Y^{ijkl}$ of $\hat Y$. Let $\{\mathcal D_1,\dots,\mathcal D_n,\mathcal D_F\}$ be the (local) Picard-Fuchs system of differential operators for period integrals of $\hat Y$. The last operator $\mathcal D_F$ is distinguished, as we take it to correspond to the complex structure variable $z_F$. 

Now, in computing fourpoint functions of $\hat Y$, we note that the system $\{\mathcal D_1,\dots,\mathcal D_n,\mathcal D_F\}$ is not sufficient. The reason for this is that these operators really correspond to the mirror of the compact threefold $\bar X$, and therefore we need more relations to compute the $Y^{ijkl}$. Our solution is to use an extended Picard-Fuchs system, as considered in \cite{FJ}. 

In order to fix a choice of extended Picard-Fuchs system, we reason as follows. From the above conjecture, classical  quadruple intersection numbers on 
$\hat X$ are given by the formula:
\begin{eqnarray}
\label{intnumbs}
	\langle C_a,C_b,C_c,C_d\rangle= \int_{\bar X}\frac{J_aJ_bJ_cJ_d}{c_1(K_{\bar X})}.
\end{eqnarray}
where $C_a \in H_2(\bar X,\Z)$ and $c_1$ is the first Chern class.  
Since $c_{1}(K_{\bar X})=-2J_{F}$, we can easily see,
\begin{eqnarray}
\label{intnumbs}
	\langle C_a,C_b,C_c,C_F\rangle= -\frac{1}{2}
\int_{\bar X}J_aJ_bJ_c,
\label{in1}
\end{eqnarray}
with no ambiguity. However, if all of the $J_{a},J_{b},J_{c},J_{d}$ are induced from 
$H^{1,1}(S,\Z)$, $\int_{\bar X}\frac{J_aJ_bJ_cJ_d}{-2J_{F}}$ cannot be computed, and we therefore have free moduli parameters. 
In this paper, we set all of these free moduli parameters to $0$, i.e., we 
set
\begin{equation}
\langle C_a,C_b,C_c,C_d\rangle= 
\int_{\bar X}\frac{J_aJ_bJ_cJ_d}{-2J_{F}}=0, 
\label{in2}
\end{equation} 
if all of $J_{a},J_{b},J_{c},J_{d}$ are induced from 
$H^{1,1}(S,\Z)$. This choice is geometrically natural since 
$J_{a}J_{b}J_{c}=0$ in $H^{*}(\bar X.\C)$ if  all of $J_{a},J_{b},J_{c}$
are induced from $H^{1,1}(S,\Z)$; furthermore, it is compatible with our choice of  moduli in the $F_{2}$ case.  

The formulas (\ref{in1}) and (\ref{in2}) completely 
fix the fractional intersection theory of $\hat X$.  
Now consider the Picard-Fuchs operators $\mathcal D_i$ to be formal polynomials in the noncommutative variables $z_i, \theta_i$, where $\theta_i=z_i \partial/\partial z_i$, and define limiting relations by the formula $R_i=\lim_{z\rightarrow 0}\mathcal D_i$. Then it is easy to show that the intersection theory defined by eqn.(\ref{in1}) and eqn. (\ref{in2}) coincides with that of the commutative ring 
\begin{eqnarray}
H^*_{ext}(\hat X,\C)=	\frac{\C[\theta_1,\dots,\theta_m,\theta_F]}{\langle R_1,\dots, R_n, \theta_F R_F   \rangle}.
\end{eqnarray}
Hence, we should choose $\{\mathcal D_1,\dots,\mathcal D_n,\theta_F \mathcal D_F\}$ as our extended Picard-Fuchs system on $\hat Y$.

With this extended system in hand, we can solve for the four point functions, but we need to make one more assumption. We assume the existence of $n$ point functions, $n=1\dots5$, which are symmetric tensors satisfying 
\bigskip
\newline
\indent (i) Griffiths transversality: the $n$ point functions vanish for $n \le 3$; 
\bigskip
\newline
\indent (ii) integrability: $2Y^{ijklm}=\theta_i Y_{jklm}+\theta_j Y_{iklm}+\theta_k Y_{ijlm}+\theta_l Y_{ijkm}+\theta_m Y_{ijkl}$,
\bigskip
\newline
\indent (iii) relations among the $n$ point functions are determined by the extended Picard-Fuchs system.
\newline

We give here a brief justification for the existence of these. This is not a proof, but is merely meant to indicate why one might expect to find $n$ point functions with the above properties. The key observation is that the extended Picard-Fuchs system $\{\mathcal D_1,\dots, \mathcal D_n, \theta_F \mathcal D_F\}$ is actually the (ordinary) Picard-Fuchs system of the toric variety $\hat X_1=\mathcal O(-\frac{1}{2}K)\oplus \mathcal O(K)\rightarrow \p(\mathcal O_S \oplus \mathcal O_S\oplus  K_S)$, where we take a section of the positive bundle $O(-\frac{1}{2}K)$ (as was done in \cite{LLY}). To see this, consider the noncompact Calabi-Yau sixfold  $\hat X_2=\mathcal O(\frac{1}{2}K)\oplus \mathcal O(K)\rightarrow \p(\mathcal O_S \oplus \mathcal O_S\oplus  K_S)$, and let $\hat Y_2$ be the mirror of $\hat X_2$, in the sense of \cite{HV}. Then we have $\hat Y_2=\{(y_1,\dots,y_5)\in (\C^*)^5: f(z_1 \dots ,z_m,z_F,y_i)=0\}$ for some $f$, and this allows us to define a meromorphic $(4,0)$ form on $\hat Y_2$ as 
\begin{eqnarray}
\hat \Omega_2=\Res_{f=0}\Big(	\prod_{i=1}^5 \frac{dy_i}{y_i}e^{-f}\Big).
\end{eqnarray}
We recall briefly the original construction of Hori-Vafa \cite{HV}. Our computation of $\hat Y_2$ may be confusing, since in the process of taking the mirror manifold, the dimension has been reduced by 2. However, this was in fact a peculiarity of their original construction. For example, the mirror of $\mathcal O(-1) \oplus \mathcal O(-1)\rightarrow \p^1$ was described in the earlier works of mirror symmetry as the hypersurface $f_1=1+y_1+y_2+zy_1y_2=0$, which is complex dimension 1. Only in slightly more recent literature do we find this equation modified to $f_2=uv+1+y_1+y_2+zy_1y_2=0$, and this is done mainly with the motivation of keeping the dimensions consistent on both sides of mirror symmetry. However, the period integrals corresponding to $f_1=0$ and $f_2=0$ are the same, so we are free to consider $\hat Y_2$ as a complex fourfold.   

Then, again using \cite{HV}, we can produce a $(4,0)$ form on $\hat Y_1$, the mirror to $\hat X_1$, by the formula $\hat \Omega_1=\theta_F \hat \Omega_2$. The derivative converts noncompact period integrals into compact ones, as follows. Recall that the result of Hori-Vafa \cite{HV} established the equation
\begin{eqnarray}
	\Pi_{compact}= \frac{\partial}{\partial t}\Pi_{non-compact}.
\end{eqnarray}
For example, this formula is well known in terms of the relationship between the period integrals of $\mathcal O(-5)\rightarrow \p^4$ and the period integrals of the quintic, which is a zero section of the bundle $\mathcal O(5)\rightarrow \p^4$.

 It is then straightforward to check that the period integrals given by $\hat \Omega_1$ are annihilated by the extended Picard-Fuchs system $\{\mathcal D_1\dots \mathcal D_n, \theta_F\mathcal D_F\}$. Thus, we have found a meromorphic form for $\hat Y_1$, which is related to the $n$ point functions as 
\begin{eqnarray}
	Y^{i_1\dots i_n}=\int_{\hat Y_1} \hat \Omega_1 \wedge \partial_{i_1}\dots \partial_{i_n} \hat \Omega_1.
\end{eqnarray}

We then move forward under the assumption of the existence of $n$ point functions. We are now able to solve for the four point functions $Y_{ijkl}$ of $\hat Y$ completely by imposing the condition that constant 
part of $Y_{ijkl}$ should coincide with $\langle C_{i}, C_{j}, C_{k}, C_{l}
\rangle$. 
Up to this point, we have completed steps $1)$ and $2)$ of the outline. Next, we transform the functions $Y_{ijkl}$ to the $A$ model via the inverse mirror map. Recall that the mirror map is given by the basis $\{t_1,...,t_m,t_F\}$ of logarithmic solutions of the Picard-Fuchs system; with this, and the knowledge that the $Y_{ijkl}$ are rank 4 tensors, we can compute the $C_{ijkl}$, which are fourpoint functions for $\hat X$. The only thing remaining is to compute $\mathcal F$ in the threefold limit, and from \cite{M}, this is done via the large fiber limit $\lim_{t_F\rightarrow - \infty}C_{ijkF}$. We now clarify the relationship between the three point functions of $K_{X}$ and $\lim_{t_F\rightarrow - \infty}C_{ijkF}$. Taking the large fiber limit corresponds to picking 
up the $d_{F}=0$ part of the fourpoint function $C_{ijkF}$. Therefore, we are led 
to consider 
\begin{equation}
\sum_{\vec{d}}q^{(\vec{d},0)}
\int_{[\overline{M}_{0}({\bar X},(\vec{d},0))]_{vir.}}
\frac{c_{top}(R^{1}\pi_{*}ev^{*}K_{\bar X})}
{c_{top}(R^{0}\pi_{*}ev^{*}K_{\bar X})}ev_{1}^{*}(J_{i})ev_{2}^{*}(J_{j})ev_{3}^{*}(J_{k})ev_{4}^{*}(J_{F}).
\label{lim}
\end{equation} 
We note that 
$\varphi\in\overline{M}_{0}({\bar X},(\vec{d},0))$ is nothing but 
the constant map along the fiber direction.
Therefore, we can regard $ev_{4}^{*}(J_{F})$ as $J_{F}$.
We can also see that  $K_{\bar X}$ is trivial along the
$X$ direction, and  we have 
\begin{equation}
c_{top}(R^{1}\pi_{*}ev^{*}K_{\bar X})=1,\;\; 
c_{top}(R^{0}\pi_{*}ev^{*}K_{\bar X})=c_{1}(K_{\bar X})=-2J_{F}. 
\end{equation}
Hence we obtain the following equality: 
\begin{eqnarray}
&&\sum_{\vec{d}}q^{(\vec{d},0)}
\int_{[\overline{M}_{0}({\bar X},(\vec{d},0))]_{vir.}}
\frac{c_{top}(R^{1}\pi_{*}ev^{*}K_{\bar X})}
{c_{top}(R^{0}\pi_{*}ev^{*}K_{\bar X})}ev_{1}^{*}(J_{i})ev_{2}^{*}(J_{j})ev_{3}^{*}(J_{k})ev_{4}^{*}(J_{F})\nonumber\\
&&=\sum_{\vec{d}}q^{(\vec{d},0)}
\int_{[\overline{M}_{0}({\bar X},(\vec{d},0))]_{vir.}}\frac{1}
{-2J_{F}}ev_{1}^{*}(J_{i})ev_{2}^{*}(J_{j})ev_{3}^{*}(J_{k})J_{F}\nonumber\\
&&=-\frac{1}{2}\sum_{\vec{d}}q^{(\vec{d},0)}
\int_{[\overline{M}_{0}({\bar X},(\vec{d},0))]_{vir.}}
ev_{1}^{*}(J_{i})ev_{2}^{*}(J_{j})ev_{3}^{*}(J_{k}).
\label{lim2}
\end{eqnarray}  
by formal reduction. Notice that if we assume that $J_{i}, J_{j}, J_{k}$
are all induced from $H^{1,1}(X,\Z)$, the constant term of 
(\ref{lim2}) vanishes. Therefore, constant term of the last line of (\ref{lim2})
vanishes. At this stage, we assume that image curve 
$C:=\varphi(\mathbb{P}^{1})$
is rigid along the fiber direction of  $\bar{X}=\mathbb{P}
({\cal O}_{S}\oplus K_{S})$.
Since $C$ is contained in S, the normal bundle $N_{C\backslash \bar{X}}$ is 
given as follows:
\begin{equation}
N_{C\backslash \bar{X}}\simeq N_{C\backslash S}\oplus K_{S}
\otimes {\cal O}_{\mathbb{P}}(1).
\label{normal}
\end{equation} 
Under the above assumption, we have to insert 
$c_{top}(R^{1}\pi_{*}ev^{*}(K_{S}\otimes {\cal O}_{\mathbb{P}}(1)))$
in reducing $[\overline{M}_{0}({\bar X},(\vec{d},0))]_{vir.}$ 
into $[\overline{M}_{0}(S,\vec{d})]_{vir.}$ since 
$c_{top}(R^{0}\pi_{*}ev^{*}(K_{S}\otimes {\cal O}_{\mathbb{P}}(1)))=1$. 
Moreover, since external 
operator insertions come only from the cohomology class of $S$, 
we can neglect the $\otimes {\cal O}_{\mathbb{P}}(1)$ part from the topological 
selection rule.  Therefore, we can rewrite the last line of (\ref{lim2}) 
as follows:
\begin{eqnarray}
&&-\frac{1}{2}\sum_{\vec{d}\neq 0}q^{(\vec{d},0)}
\int_{[\overline{M}_{0}({\bar X},(\vec{d},0))]_{vir.}}
ev_{1}^{*}(J_{i})ev_{2}^{*}(J_{j})ev_{3}^{*}(J_{k})\nonumber\\
&&=-\frac{1}{2}\sum_{\vec{d}\neq 0}q^{(\vec{d},0)}
\int_{[\overline{M}_{0}(S,\vec{d})]_{vir.}}
c_{top}(R^{1}\pi_{*}ev^{*}(K_{S}))
ev_{1}^{*}(J_{i})ev_{2}^{*}(J_{j})ev_{3}^{*}(J_{k}).
\label{lim3}
\end{eqnarray}
Since $c_{top}(R^{0}\pi_{*}ev^{*}(K_{S}))=1$,
we can see that the last line of (\ref{lim3}) coincides with the formula
(\ref{st}) up to the factor $-\frac{1}{2}$ (neglecting constant terms).
Therefore, we have obtained the following equation under the assumption that
all the image curves are rigid along the fiber direction of $\bar{X}=\mathbb{P}
({\cal O}_{S}\oplus K_{S})$:
\begin{equation}
\lim_{t_F\rightarrow - \infty}C_{ijkF}=\frac{-1}{2}\partial_{i}\partial_{j}\partial_{k}{\mathcal F}_{inst.}.
\label{predef}
\end{equation}  
   Then equation (\ref{predef}) gives a defining equation for the prepotential $\mathcal F$, up to polynomial terms! We emphasize that while the usual local mirror symmetry Picard-Fuchs system is only able to identify a single linear combination of prepotential derivatives, the above formula completely fixes $\mathcal F$ up to the polynomial ambiguity. This is particularly handy when the number of K\"ahler parameters of $X=K_S$ becomes large. 
   
Of course, if $S=F_{2}$,
the above assumption breaks down for the $(0,1)$ curve mentioned in the 
previous section. We discuss here 
this situation in detail. The toric construction of 
$\mathbb{P}({\mathcal O}\oplus{\mathcal O}_{K_{F_2}})$ is obtained by adding a sixth 
variable $x_{6}$ to the toric construction of $K_{F_{2}}$ in the previous section.
\begin{eqnarray}
&&(x_{1},x_{2},x_{3},x_{4},x_{5},x_{6})
\sim(x_{1},x_{2},\mu x_{3},\mu x_{4},\mu^{-2}x_{5},x_{6})
\sim(\lambda x_{1},\lambda x_{2},x_{3},\lambda^{-2}x_{4},x_{5},x_{6})
\nonumber\\
&&\sim(x_{1},x_{2},x_{3},x_{4},\nu x_{5},\nu x_{6}).
\label{pkf2}
\end{eqnarray}   
Let $J_{u},J_{v},J_{w}$  be the K\"ahler forms associated with the actions
$\mu, \lambda, \nu$ respectively. These forms are 
generators of the classical cohomology ring of 
$\mathbb{P}({\mathcal O}\oplus{\mathcal O}_{K_{F_2}})$. 
Relations of the classical cohomology 
ring are given by,
\begin{equation}  
J_{u}^{2}=2J_{u}J_{v},\;\;J_{v}^{2}=0,\;\;J_{w}^{2}=2J_{w}J_{u}.
\label{relpkf2}
\end{equation}
This space is a $\mathbb{P}^{1}$ fibration of $F_{2}$, but it can also be 
regarded as an $F_{2}$ fibration of $\mathbb{P}^{1}$. We denote the $F_2$ fiber whose cohomology ring is 
generated by $J_{u}$ and $J_{w}$ by $F^{f}_{2}$. We have relations:
\begin{equation}
J_{u}^{2}=0,\;\;J_{w}^{2}=2J_{w}J_{u}.
\end{equation}
Then degree $((0,1),0)$ map of $\mathbb{P}({\mathcal O}\oplus{\mathcal O}_{K_{F_2}})$
is given as follows:
\begin{equation}
\varphi(s,t):=(a_{1}s+a_{2}t,b_{1}s+b_{2}t,c,0,e,f).
\label{phol}
\end{equation}
Therefore, we can see that moduli space of $\varphi$ is compactified 
into $\mathbb{P}^{3}\times\mathbb{P}^{1}$ by considering 
three $C^{*}$ action. The second $\mathbb{P}^{1}$ is Poincare dual 
of $k_{u}$ in $F^{f}_{2}$.
With this setting, we compute\\ 
$\langle {\cal O}_{J_{v}}(z_{1}){\cal O}_{J_{v}}(z_{z}){\cal O}_{J_{v}}(z_{3}){\cal O}_{J_{F}}(z_{4})
\rangle_{((0,1),0)}$. In the same way as the first part of the 
 previous computation, we can derive,
\begin{eqnarray}
&&\langle {\cal O}_{J_{v}}(z_{1}){\cal O}_{J_{v}}(z_{z}){\cal O}_{J_{v}}(z_{3}){\cal O}_{J_{F}}(z_{4})
\rangle_{((0,1),0)}
=-\frac{1}{2}
\int_{[\overline{M}_{0}(\mathbb{P}({\cal O}_{K_{F_{2}}}\oplus{\cal O}),((0,1),0))]_{vir.}}ev_{1}^{*}(J_{v})
ev_{2}^{*}(J_{v})ev_{3}^{*}(J_{v}).\nonumber\\
&&
\label{ex}
\end{eqnarray}
The obstructed normal bundle of the image curve is generated by $x_{4}$ and 
it is isomorphic to ${\cal O}_{F_{2}}(-2J_{v}+J_{u})$. This generates the same 
virtual fundamental class $J_{u}$ as the discussion in the previous section. 
By using the above compactification of the moduli space, we can proceed 
as follows:
\begin{eqnarray}
\langle {\cal O}_{J_{v}}(z_{1}){\cal O}_{J_{v}}(z_{z}){\cal O}_{J_{v}}(z_{3}){\cal O}_{J_{F}}(z_{4})
\rangle_{((0,1),0)}
&=&-\frac{1}{2}
\int_{\overline{M}_{0}(\mathbb{P}({\cal O}_{K_{F_{2}}}\oplus{\cal O}),((0,1),0))}J_{u}ev_{1}^{*}(J_{v})
ev_{2}^{*}(J_{v})ev_{3}^{*}(J_{v})\nonumber\\
&=&-\frac{1}{2}
\int_{\mathbb{P}^{3}}H^{3}\int_{\mathbb{P}^{1}}J_{u},\nonumber\\
&=&-\frac{1}{2}\int_{\mathbb{P}^{1}}J_{u},
\end{eqnarray}
where $H$ is the hyperplane class of $\mathbb{P}^{3}$.
At this stage, we have to remember the fact that this $\mathbb{P}^{1}$ is
identified with $PD(J_{u})$ in $F^{f}_{2}$. Hence we have,
\begin{equation}
\langle {\cal O}_{J_{v}}(z_{1}){\cal O}_{J_{v}}(z_{z}){\cal O}_{J_{v}}(z_{3}){\cal O}_{J_{F}}(z_{4})
\rangle_{((0,1),0)}
=-\frac{1}{2}\int_{\mathbb{P}^{1}}J_{u}=
-\frac{1}{2}\int_{F_{2}^{f}}J_{u}^{2}=0.
\end{equation}
From this result, we conclude that this curve cannot be detected 
from the local fourfold computation. 

Next, suppose that $\dim H_4(X,\Z)=0$. First we discuss  $\dim H_2(X,\Z)=1$ 
cases. In this paper, we treat ${\bar X}=\mathbb{P}({\cal O}\oplus{\cal O}(-1)
\oplus{\cal O}(-1)),\mathbb{P}({\cal O}\oplus{\cal O}\oplus{\cal O}(-2))$ 
which are compactifications of $X={\cal O}(-1)\oplus{\cal O}(-1)
\rightarrow \mathbb{P}^{1},{\cal O}\oplus{\cal O}(-2)\rightarrow 
\mathbb{P}^{1}$. Here, we denote by $H$ the hyperplane class of the base 
$\mathbb{P}^{1}$. We also denote   
$c_{1}({\cal O}_{\mathbb{P}}(1))$, which is a 
generator of the cohomology class of fiber direction, by $J_{F}$. 
Since $c_{1}(K_{\bar X})=-3J_{F}$, we can compute the large fiber limit 
of $C_{HHHF}$ in the same way as the first half of the computation of the
$X=K_{S}$ case:
\begin{equation} 
-\frac{1}{3}\sum_{d>0}q^{(d,0)}
\int_{[\overline{M}_{0}({\bar X},(d,0))]_{vir.}}
ev_{1}^{*}(H)ev_{2}^{*}(H)ev_{3}^{*}(H)
\end{equation}
The remaining computations depend on the structure of the fibers; we discuss
the ${\bar X}=\mathbb{P}({\cal O}\oplus{\cal O}(-1)
\oplus{\cal O}(-1))$ case first. In this case, the image curve is rigid in 
the fiber direction, and we only have to insert $c_{top}(\pi_{*}ev^{*}
({\cal O}(-1)\oplus{\cal O}(-1)))$ to reduce 
$\overline{M}_{0}({\bar X},(d,0))$ into 
$\overline{M}_{0}({\mathbb{P}^{1}},d)$. Therefore, we obtain the 
following formula:
\begin{eqnarray}
\lim_{t_{F}\rightarrow\infty}C_{HHHF}&=&-\frac{1}{3}\sum_{d>0}q^{(d,0)}
\int_{[\overline{M}_{0}({\bar X},(d,0))]_{vir.}}
ev_{1}^{*}(H)ev_{2}^{*}(H)ev_{3}^{*}(H)\nonumber\\
&=&-\frac{1}{3}\sum_{d>0}q^d
\int_{\overline{M}_{0}(\mathbb{P}^{1},d)}c_{top}(\pi_{*}ev^{*}
({\cal O}(-1)\oplus{\cal O}(-1)))
ev_{1}^{*}(H)ev_{2}^{*}(H)ev_{3}^{*}(H)\nonumber\\
&=&-\frac{1}{3}\cdot\frac{q}{1-q},
\label{-1}
\end{eqnarray}
as is well known from the result of Aspinwall and Morrison. 

Next, we discuss the ${\bar X}=\mathbb{P}({\cal O}\oplus{\cal O}
\oplus{\cal O}(-2))$ case. In this case, the image curve is not rigid in the 
fiber direction, and $\overline{M}_{0}({\bar X},(d,0))$ turns out to be 
$\mathbb{P}^{1}\times\overline{M}_{0}({\mathbb{P}^{1}},d)$, where 
the left  
$\mathbb{P}^{1}$ is contained in the fiber $\mathbb{P}^2$ of $\bar X$.
Therefore, we have to insert $c_{top}(\pi_{*}ev^{*}
({\cal O}(-2)\otimes{\cal O}_{\mathbb{P}}(1)))=
\sum_{j=0}^{2d-1}c_{j}(\pi_{*}ev^{*}({\cal O}(-2)))
J_{F}^{2d-1-j}$ in order to reduce 
$\overline{M}_{0}({\bar X},(d,0))$ into $\mathbb{P}^{1}\times 
\overline{M}_{0}({\mathbb{P}^{1}},d)$.
With these considerations, we obtain the following result:
\begin{eqnarray}
\lim_{t_{F}\rightarrow\infty}C_{HHHF}&=&-\frac{1}{3}\sum_{d>0}q^{(d,0)}
\int_{[\overline{M}_{0}({\bar X},(d,0))]_{vir.}}
ev_{1}^{*}(H)ev_{2}^{*}(H)ev_{3}^{*}(H)\nonumber\\
&=&-\frac{1}{3}\sum_{d>0}q^d
\int_{\mathbb{P}^{1}\times\overline{M}_{0}(\mathbb{P}^{1},d)}
c_{2d-2}(\pi_{*}ev^{*}
({\cal O}(-2)))\cdot J_{F}\cdot
ev_{1}^{*}(H)ev_{2}^{*}(H)ev_{3}^{*}(H)\nonumber\\
&=&-\frac{1}{3}\sum_{d>0}q^d
\int_{\overline{M}_{0}(\mathbb{P}^{1},d)}
c_{2d-2}(\pi_{*}ev^{*}
({\cal O}(-2)))
ev_{1}^{*}(H)ev_{2}^{*}(H)ev_{3}^{*}(H)\nonumber\\
&=&-\frac{1}{3}\cdot\frac{q}{1-q},
\label{-2}
\end{eqnarray}
which follows from the localization computation.

We then turn to the  $\dim H_2(X,\Z)> 1$ case. 
Here, we briefly discuss the schematic procedure needed for the mirror computation.
In this case, 
 we must add the following steps to those used for $K_S$:
\bigskip
\newline
\indent 0) Compactify the moduli space of all curve classes $C$ such that $N_{C/X}\cong \mathcal O \oplus \mathcal O(-2)$, 
\bigskip
\newline
\indent $\frac{1}{2}$) Flop the resulting space to a canonical bundle model,  
\bigskip
\newline
\indent $3 \frac{1}{3}$) Reverse the flop transition of step $\frac{1}{2}$,
\bigskip
\newline
\indent $3 \frac{2}{3}$) Decompactify the compactified moduli spaces of step 0.
\newline 

\begin{figure}[t]
\label{cptification}
\centering
\input{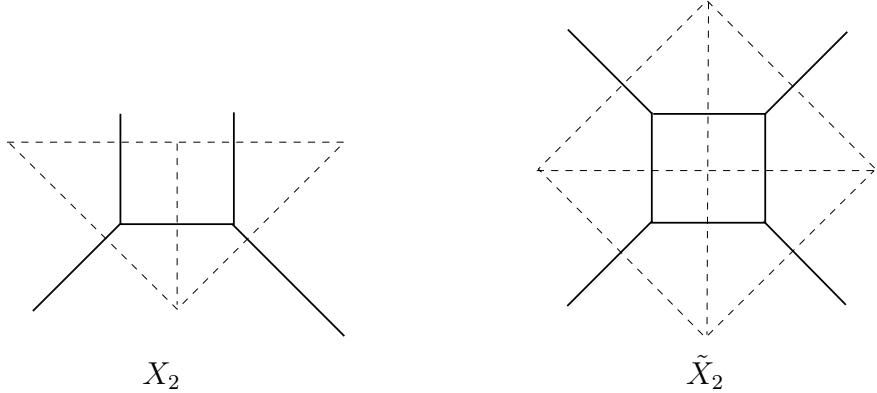}
\caption{Toric diagram of the partial compactification on $\mathcal O \oplus \mathcal O(-2)\longrightarrow \p^1$.}
\label{p1completionsecond}
\end{figure}

Here we are assuming that the `compactified' model we get after step 0 admits a flop to a canonical bundle type space. This certainly holds true in all the examples we consider, and probably has a reasonably broad range of validity. Then the only step in the above which is not self-explanatory is number 0, since there are clearly a variety of compactifications available, and the result varies demonstrably with the choice. Our approach is to use a compactification such that the outcome is consistent with topological vertex calculations \cite{I}. This compactification was first considered in \cite{DFG}, and the basic example of it is depicted in figure \ref{p1completionsecond}.

We mention here one extra subtlety which comes along with the use of this compactification scheme in computing the prepotential. We are using this compactification because, as mentioned, we cannot see the presence of $-2$ curves from usual mirror symmetry calculations. However, using this compactification, we actually find the result that the $-2$ curves are overcounted by a factor of 2. The reason for this is as follows. For any toric graph containing Figure \ref{cptification} as a subgraph, the instanton number of the relevant curve comes out to be $-2$. This is because Figure \ref{cptification} is the graph of $K_{F_0}$, and the Gromov-Witten invariant of each curve of $K_{F_0}$ is $-2$. Therefore, at the last step we should divide the resulting Gromov-Witten invariant for the compactified curves by 2. 

Then, by working through steps 0-4 for the $\dim H_4(X,\Z)=0$ case, we are able to calculate the correct prepotential in a number of examples. In addition, we carry out the computation for one example whose prepotential has not been worked out elsewhere, and find a result which might have been guessed from the findings of \cite{I}.

\section{Local Calabi-Yau fourfolds.}

We now turn our attention to the mirror symmetry construction of the prepotential for noncompact Calabi-Yau threefolds.

\subsection{Fourfold compactifications of local threefolds.}

We begin this section by offering some motivation on the utility of local Calabi-Yau fourfolds. We will demonstrate that local fourfolds are one of the more natural objects one might consider in cases where ordinary local mirror symmetry for threefolds breaks down. To this end, consider the space $X=\mathcal O \oplus \mathcal O(-2)\longrightarrow \p^1$. This can be realized as a symplectic quotient
\begin{eqnarray}
X=\{(z_1,\dots,z_4)\in \C^4-Z:-2|z_1|^2+|z_2|^2+|z_3|^2=r\}/S^1.	
\end{eqnarray}
Here $Z=\{z_2=z_3=0\}$ is the exceptional locus, $r \in \mathbb R^+$ and 
\begin{eqnarray}
S^1:(z_1,\dots,z_4)\longrightarrow (e^{-2i\theta}z_1,e^{i\theta}z_2,e^{i\theta}z_3,z_4), \ \ \theta \in S^1.
\end{eqnarray}
Note that the vector $(-2 \ 1 \ 1 \ 0)$ completely specifies the geometry of $X$.

The usual constructions of local mirror symmetry \cite{H} fail for this case, because the Picard-Fuchs operator is only of order 2, and its solutions are spanned by ${1,t}$ where $t$ is the mirror map. This constitutes a failure of mirror symmetry exactly because there is one holomorphic curve in $X$, and this curve is not counted, as we would like. Recently, one remedy for this was offered in \cite{FJ}, where an extended Picard-Fuchs operator was constructed. Here, we will take a different approach.

One of the reasons for the problem of the uncounted curve is that $\p^1\hookrightarrow X$ has a noncompact deformation space $\C$. Hence, we should be able to recover the curve information by compactifying this deformation space; the simplest choice for such an operation is the projective closure $\bar X$, which is the compact toric manifold given by the vectors
\begin{eqnarray}
\begin{pmatrix}
-2 & 1 & 1 & 0 & 0  \\ 1 & 0 & 0 & 1 & 1  
\end{pmatrix}.
\end{eqnarray}
We have $\bar{X} \cong \p(\mathcal O\oplus \mathcal O \oplus \mathcal O(-2))\longrightarrow \p^1$.
Notice that this is a $\p^2$ fibration over $X$. Unfortunately, this new space is not Calabi-Yau, but there is a natural local CY fourfold associated to it:
\begin{eqnarray}
K_{\bar X}=\{-2|z_1|^2+|z_2|^2+|z_3|^2=r,-3|z_0|^2+|z_1|^2+|z_4|^2+|z_5|^2=r_F\}/(S^1)^2.		
\end{eqnarray}
This is, of course, just the local CY given by 
\begin{eqnarray}
\begin{pmatrix}
0& -2 & 1 & 1 & 0 & 0  \\ -3 & 1 & 0 & 0 & 1 & 1  
\end{pmatrix}
\end{eqnarray}
and is the canonical bundle over $\bar X$. Now, recall \cite{H}\cite{FJ} that local mirror symmetry on a space $X$ is incomplete (i.e., the prepotential cannot be reconstructed from solutions of the PF operators) exactly when $\dim H_2(X)\ne \dim H_4(X)$. In the case of $\mathcal O \oplus \mathcal O(-2)\longrightarrow \p^1$, there are no four cycles at all, which translates into a lack of predictive power of the instanton expansion via mirror symmetry. The new space $K_{\bar X}$ has two four cycles, and moreover the deformation space of the base curve has been compactified, which indicates that this geometry should have the instanton numbers that were lacking on $X$. 

On any space $X=K_S$, the canonical bundle over a surface $S$, we can give a general description of this procedure via charge vectors. First, write the charge vectors  of $X$
\begin{eqnarray}
\begin{pmatrix}
-l_0^1 & l_1^1 & \dots & l_n^1  \\ \vdots & & & \vdots \\  -l_0^{n-2} & l_1^{n-2} & \dots & l_n^{n-2}
\end{pmatrix}	
\end{eqnarray}
where we take the convention that $l_0^i \ge 0 \ \forall i$. This means that, if $[C_i]$ is the curve class associated to the vector $l^i$, then the canonical bundle of $S$ is $\sum_i l^i_0 [C_i]$. Then we define the associated noncompact Calabi-Yau fourfold to be 
\begin{eqnarray}
\begin{pmatrix}
0 & -l_0^1 & l_1^1 & \dots & l_n^1 & 0  \\ \vdots & \vdots & & & \vdots & \vdots \\ 0 & -l_0^{n-2} & l_1^{n-2} & \dots & l_n^{n-2} & 0 \\ -2 & 1 & 0 & \dots & 0 & 1.
\end{pmatrix}		
\end{eqnarray}
which is nothing but the canonical bundle over $\p(\mathcal O_S \oplus K_S)$.
Note that, while we can associate a noncompact fourfold to any geometry of type $K_S$, we only expect that the Picard-Fuchs system on the fourfold has new information about curves in $S$ if $\dim H_2(S)\ne \dim H_4(S)$. 

We now move on to discuss the methods of analyzing local fourfold geometries.

\subsection{Periods of Local Fourfolds.}

Here, we will briefly describe relevant geometric quantities of fourfolds in terms of Picard-Fuchs solutions. See \cite{M} for a similar discussion for compact fourfolds. 

We assume that we begin with a noncompact Calabi-Yau threefold $X_0$, and let $B^3$ be the projective closure of $X_0$. Then the fourfolds we will use are all of the type $X=K_{B^3}$, where $K_{B^3}$ is the canonical bundle over $B^3$. This is specified by a set of vertices $\{\nu_1,\dots,\nu_n\}\subset \Z^4$. Choose a basis of relation vectors $\{l^1,\dots,l^m\}$ satisfying $\sum_{k}l_i^kv_k=0 \ \forall i$, and let $C_1,\dots,C_m$ be the corresponding basis of $H_2(X,\Z)$. Then we take $\{J_1,\dots,J_m\}$ as a basis of $H^{1,1}(X,\C)$, where $\int_{C_i}J_j=\delta_{ij}$. Next, take $D_1,\dots,D_k$ to be the basis of $H_4(X,\Z)$ corresponding to the columns of the intersection matrix (i.e. $D_i\cap C_j=l^j_i$). Note that while every row vector of the charge matrix gives us a 2 cycle, not every column of the charge matrix corresponds to a compact 4 cycle. A particular column will give a compact four cycle if its corresponding vertex is an interior point of the convex hull of $\{\nu_1,\dots,\nu_n\}$. We can then define a dual basis of four forms $\sum_{j,k}c_b^{jk}J_j\wedge J_k$ by the equation
\begin{eqnarray}
	\int_{D_a}\sum_{j,k}c_b^{jk}J_j\wedge J_k=\delta_{ab}.
\end{eqnarray}
Finally, note that there is a single 6 form which satisfies
\begin{eqnarray}
	\int_{B^3}\sum_{ijk}a^{ijk}J_i\wedge J_j\wedge J_k =1.
\end{eqnarray}

Now let $Y$ be the mirror of $X$. Then using the lattice vectors $\{l^1,\dots,l^m\}$, we can immediately write down a Picard-Fuchs system of differential operators $\{\mathcal D_1, \dots, \mathcal D_j \}$ such that the solution space of the differential equations is the same as the period integrals of $Y$. The generating function of solutions for this system is 
\begin{eqnarray}
	\omega=\sum_{n\ge 0}\prod_j\big(\Gamma(1+\sum_{i}l^j_i(n_i+\rho_i))\big)^{-1}z^{n+\rho}.
\end{eqnarray}
 Then, using the above bases of cohomology on $X$, we can describe the solution space of $\{\mathcal D_1, \dots, \mathcal D_j \}$ as follows. Let $\Pi_{ij}=\partial_{\rho_i}{\partial}_{\rho_j}\omega|_{\rho=0}$. The solution space becomes
\begin{eqnarray}
	\Big(1,\Pi_1,\dots,\Pi_m,\sum_{j,k}c_1^{jk}\Pi_{jk},\dots,\sum_{j,k}c_m^{jk}\Pi_{jk},\sum_{i,j,k}d^{ijk}\Pi_{ijk}\Big).
\end{eqnarray}
Here, the $c_a^{jk}$ are the same as in the $X$ case, and 
\begin{eqnarray}
	d^{ijk}=\int_{B^3}J_i\wedge J_j \wedge J_k.
\end{eqnarray}

With this data, we can construct the fourpoint functions of $Y$. Let $\eta^{ab}$ be the intersection matrix of four cycles on $X$, $\eta^{ab}=D_a \cdot D_b$. Also, set $\Pi_k=t_k$ and $\sum_{j,k}c_a^{jk}\Pi_{jk}=W_a$. Then the threepoint functions are defined by 
\begin{eqnarray}
	Y_{\alpha \beta \gamma}=\partial_{t_{\alpha}} \partial_{t_{\beta}} W_{\gamma}.
\end{eqnarray}
Note that while the solutions $W_a$ of the Picard-Fuchs system have double logarithmic singularities, the threepoint functions are holomorphic in $z$. The fourpoint functions are then 
\begin{eqnarray}
	Y_{\alpha \beta \gamma \delta}=\sum_{a,b}Y_{\alpha \beta a}\eta^{ab}Y_{b \gamma \delta},
\end{eqnarray}
and these are also holomorphic in $z$. 

Finally, there is one more fact about these fourpoint functions which we will make heavy use of \cite{M}. Note that for the compactification $B^3 \rightarrow X_0$, with $X_0$ the given noncompact Calabi-Yau threefold, the number of K\"ahler parameters has increased by 1. Let $t_{fiber}=t_m$ be the K\"ahler parameter corresponding to the compactification $B^3\rightarrow X_0$. With the above conventions, we therefore have that $\{C_1,\dots, C_{m-1}\}$ is a basis of $H_2(X_0,\Z)$. If we take $C^{inst.}_{abc}$ to be the instanton part of 
the Yukawa couplings for $X_0$, then we can compute the $C^{inst.}_{abc}$ from the $Y_{\alpha \beta \gamma m}$ in the following limit:
\begin{eqnarray}
   \label{threefoldlimit}
	\lim_{t_{m}\rightarrow -\infty}Y_{\alpha \beta \gamma m}=
\bigl(\frac{J_{m}}{c_{1}(B^{3})}\bigr)\cdot C^{inst.}_{\alpha \beta \gamma},
\end{eqnarray}
which follows from the result in Section 3.
In what follows, our main strategy will be to compute the fourpoint functions for $X$ and then derive the threepoint functions on $X_0$ in the above limit. Note that we must perform the above limit in $A$ model coordinates, i.e. the coordinates on the complexified K\"ahler moduli space of $X$.

\section{Some Examples.}

\subsection{Application to local $\p^1$.}

We will here apply the canonical bundle over the projective completion technique to a local $\p^1$ with normal bundle either $\mathcal O(-1)\oplus \mathcal O(-1)$ or $\mathcal O\oplus \mathcal O(-2)$. In both cases, we find that the resulting noncompact fourfold contains the instanton data in a natural way.
\begin{example}\normalfont
First, we note in greater detail why it is that one might see missing instanton information in the noncompact fourfold geometry. Consider $X=\mathcal O(-1)\oplus \mathcal O(-1)\longrightarrow \p^1$, which is determined by the vector $\begin{pmatrix}1&1&-1&-1\end{pmatrix}$. We associate to $X$ the noncompact fourfold $K_{\bar X}$, described by the vectors 
\begin{eqnarray}
\label{-1-1closure}
\begin{pmatrix}
0 & 1 & 1 & -1 & -1 & 0  \\ -3 & 0 & 0 & 1 & 1 & 1 
\end{pmatrix}.	
\end{eqnarray}
This is the canonical bundle over $\p(\mathcal O \oplus \mathcal O(-1) \oplus \mathcal O(-1))\longrightarrow \p^1$.
There is a nice graphical representation of this procedure, as illustrated in Figure \ref{p1completion}.
\begin{figure}[t]
\centering
\input{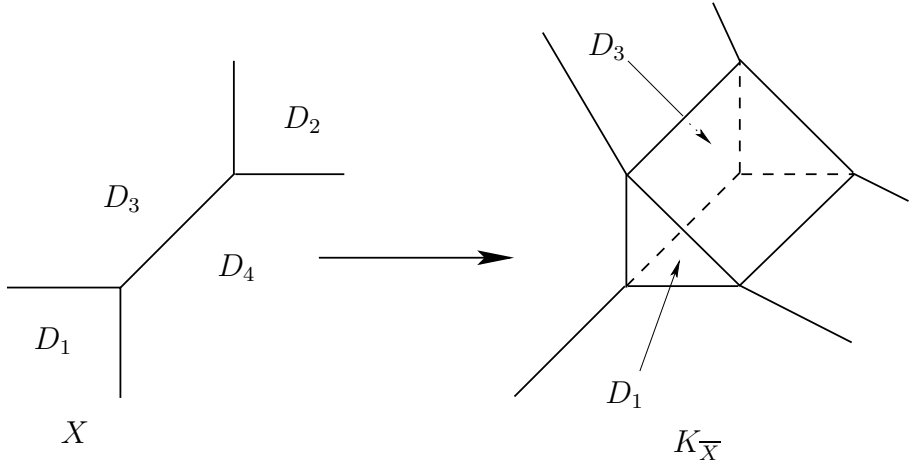}
\caption{The projective closure procedure. The external lines on the $K_{\bar X}$ picture represent the canonical bundle direction.}
\label{p1completion}
\end{figure}
By looking at this picture, we can gain an understanding about what the projective closure does for us computationally. Recall \cite{HV} that on the geometry $\mathcal O(-1)\oplus \mathcal O(-1)\longrightarrow \p^1$, we are supposed to be able to recover the instanton data by computing the `volume of the noncompact 4-cycle dual to the $\p^1$'. This is made into a sensible calculation in that paper by introducing a cutoff parameter on this 4-cycle and performing the regulated integral. Yet, from our picture here, we can see that the noncompact 4-cycle is given a finite volume; and moreover, we can find that volume simply by analyzing the period integrals on the mirror of $K_{\bar X}$. We can then recover the data originally coming from $\mathcal O(-1)\oplus \mathcal O(-1)\longrightarrow \p^1$ by taking the large fiber limit on the relevant integrals.

With that being said, we begin the computation.
Denote the mirror of $K_{\bar X}$ by $Y$. Then $Y$ is a CY fourfold which can be described by the equation
\begin{eqnarray}
\label{mirror1}
Y=\{uv+1+y_2+z_1y_4y_5/y_2+y_4+y_5+z_2/(y_4y_5)=0\}	
\end{eqnarray}
where $u,v\in \C$ and $y_i\in \C^*. $\footnote{Note that this mirror manifold is slightly different from that used in section 3. While the Picard-Fuchs operators will be the same, this description has a factor of $uv$ in front for dimensional reasons.} The Picard-Fuchs differential operators for period integrals on $Y$ are 
\begin{eqnarray}
\label{p1PF1}
\mathcal D_1&=&\theta_1^2-z_1(-\theta_1+\theta_2)(-\theta_1+\theta_2) \\ 	
\no \mathcal D_2&=&\theta_2(\theta_2-\theta_1)^2-z_2(-3\theta_2)(-3\theta_2-1)(-3\theta_2-2).
\end{eqnarray}
The Poincare polynomial is 
\begin{eqnarray}
\frac{(1-t^2)(1-t^3)}{(1-t)^2}=t^3+2t^2+2t+1	
\end{eqnarray}
which gives exactly the right number of 0,2,4 and 6 cycles, as is clear from Figure \ref{p1completion}. Corresponding to the two two cycles in the $A$ model geometry, there are two logarithmic solutions $t_1,t_2$ of the system (\ref{p1PF1}), two double logarithmic ones for the four cycles, $W_1$ and $W_2$, and of course we have a solution from the six cycle. 

Consider now  the extended system of differential operators $\{\mathcal D_1, \theta_2 \mathcal D_2\}$. The Poincare polynomial of $\{\mathcal D_1, \theta_2 \mathcal D_2\}$ is indeed such that we should expect its solutions to be of the type usually associated to a compact Calabi-Yau fourfold. We let $M$ be a fourfold with period integrals coincident with the solutions of $\{\mathcal D_1, \theta_2\mathcal D_2\}$. As explained in section 3, we can take $M$ as the mirror of $\mathcal O(H_F)\oplus \mathcal (-3H_F)\rightarrow \p(\mathcal O\oplus \mathcal O\oplus \mathcal O(-1)\oplus \mathcal O(-1))$, where $H_F$ is the class dual to the fiber curve $C_F=C_2$. Set  
\begin{eqnarray}
Y^{mn}_{(k)}&=&\int_M\Omega\wedge\nabla^m_{\delta_{z_1}}\nabla^n_{\delta_{z_2}} \Omega, \ m+n=k, \ k \in \{4,5\}, \\
\no  Y^{mn}_{(k)}&=&0, \ k \le 3
\end{eqnarray}
where $\Omega$ is the (4,0) form on $M$ and $\nabla$ is the connection on the complex structure moduli space of $M$. We can then use the extended Picard-Fuchs equations to derive relations among the $Y^{mn}_{(4)}$. To see this in the present case, note that we have exactly four equations 
\begin{eqnarray}\no
\theta_1^2\mathcal D_1f=0, \ \ \theta_2^2\mathcal D_1f=0, \\ \theta_1\theta_2\mathcal D_1f=0, \ \
\theta_2\mathcal D_2f=0,
\end{eqnarray}
and these imply the following relations for four point functions:
\begin{eqnarray}\no
	\big(-Y^{22}_{(4)}+2Y^{13}_{(4)}-Y^{04}_{(4)}\big)z_1+Y^{22}_{(4)}&=&0 \\ \big(-Y^{31}_{(4)}+2Y^{22}_{(4)}-Y^{13}_{(4)}\big)z_1+Y^{31}_{(4)}&=&0 \\ \big(-Y^{40}_{(4)}+2Y^{31}_{(4)}-Y^{22}_{(4)}\big)z_1+Y^{40}_{(4)}&=&0 \\
	27Y^{04}_{(4)}z_2+Y^{22}_{(4)}-2Y^{13}_{(4)}+Y^{04}_{(4)}&=&0.
\end{eqnarray}
Solving these relations completely determines the $Y^{mn}_{(4)}$, up to the overall multiplicative function $S=Y^{04}_{(4)}$. We can then use the PF system again (this time with one higher power of derivatives) to derive a system of partial differential equations for $S$. To see how this works, note that the assumption of the existence of $M$ made above implies a relationship between four point and five point functions:
\begin{eqnarray}
	Y^{mn}_{(5)}=\frac{1}{2}\big(m\theta_1 Y^{m-1,n}_{(4)}+n\theta_2 Y^{m,n-1}_{(4)}\big).
\end{eqnarray}
Then one could use this formula, together with a degree 5 relation (for example, $\theta_1^2\theta_2\mathcal D_1f=0$) in order to write down partial differential equations for $S$. If we solve these partial differential equations in our present case, the result is  $S^{-1}=\Delta_f=1+54z_2+54z_1z_2+729z_2^2-1458z_1z_2^2+729z_1^2z_2^2$. 
We note that $\Delta_f$ is exactly the discriminant locus of the hypersurface in eq.(\ref{mirror1}). This turns out to be the case for all the examples we consider. 

The overall normalization of the  four point functions are determined from the 
result in Section 3. We can read off the relations of classical cohomology ring 
of $\bar X$ from (\ref{p1PF1}) as follows:  
\begin{equation}
k_{1}^{2}=0,\;\;k_{2}(k_{2}-k_{1})^{2}=0.
\end{equation}
Then we obtain,
\begin{eqnarray}
&&\langle C_{2}, C_{2}, C_{2}, C_{2} \rangle
=\int_{\bar X}\frac{k_{2}^{4}}{-3k_{2}}
=-\frac{2}{3},\;\;
\langle C_{1}, C_{2}, C_{2}, C_{2} \rangle
=\int_{\bar X}\frac{k_{1}k_{2}^{3}}{-3k_{2}}
=-\frac{1}{3},\nonumber\\
&&\langle C_{1}, C_{1}, C_{2}, C_{2} \rangle=
\int_{\bar X}\frac{k_{1}^2k_{2}^{2}}{-3k_{2}}
=0,\;\;
\langle C_{1}, C_{1}, C_{1}, C_{2} \rangle=
\int_{\bar X}\frac{k_{1}^3k_{2}}{-3k_{2}}
=0,\;\;
\nonumber\\
&&\langle C_{1}, C_{1}, C_{1}, C_{1} \rangle=0,\;\;
\end{eqnarray}

So, back to the calculation. With the above, we have completely solved for the four point functions: 
\begin{eqnarray}
Y^{04}_{(4)}=-\frac{2}{3}\cdot\frac{1}{\Delta_f} \ , \ \	
Y^{13}_{(4)}=-\frac{2}{3}\cdot\frac{1-27z_1+27z_1z_2}{2\Delta_f} \ , \ \	
Y^{22}_{(4)}=-\frac{2}{3}\cdot\frac{27z_1z_2}{\Delta_f} \ , \\
\no Y^{31}_{(4)}=-\frac{2}{3}\cdot\frac{z_1(-1+27z_2+81z_1z_2)}{2(z_1-1)\Delta_f} 
\ , \ \	
Y^{40}_{(4)}=-\frac{2}{3}\cdot\frac{z_1^2(-1+54z_2+54z_1z_2)}{(z_1-1)^2\Delta_f} \ .
\end{eqnarray}
These are not terribly enlightening in this form, but we can perform a coordinate change to the $A$ model using the inverse mirror map (treating the above functions as rank 4 tensors). Let $C^{mn}_{(4)}$ be the resulting $A$ model fourpoint functions. We have, in particular,
\begin{eqnarray}
\lim_{t_2\rightarrow -\infty}C^{31}_{(4)}=-\frac{1}{3}\sum_{n\ge 1}e^{nt_1}.	
\end{eqnarray}
Here $t_1,t_2$ are the logarithmic solutions of the PF system. 
Therefore we have obtained the instanton part of 
 the prepotential for this space by the equation
\begin{eqnarray}
	\frac{d^3 \mathcal F_{inst.}}{d t^3}= -3\lim_{t_2\rightarrow -\infty} C^{31}_{(4)}.
\end{eqnarray}
\end{example}

\begin{example}\normalfont
We now present the result of applying the same procedure to $\mathcal O \oplus \mathcal O(-2)\longrightarrow \p^1$. Since this is nearly the same as the above, we give only the briefest overview. We mention, however, that the process of taking the projective completion adds \it more \normalfont information than in example 1. This is because in example 1, there was already a rigid curve which could in principle be counted through other means. Here, we have additionally compactified the deformation space of the curve, which amounts to a nontrivial addition of Gromov-Witten information.

Recall that the defining vectors are
\begin{eqnarray}
\begin{pmatrix}
0&-2 & 1 & 1 & 0 & 0  \\ -3&1 & 0 & 0 & 1 & 1  
\end{pmatrix}.
\end{eqnarray} 
The mirror geometry is
\begin{eqnarray}
Y=\{uv+1+z_2/(y_5y_6)+z_1z_2^2/(y_4y_5^2y_6^2)+y_4+y_5+y_6=0\}
\end{eqnarray}
The discriminant locus of this hypersurface is 
\begin{eqnarray}
	\Delta_f=1+54z_2+729z_2^2-2916z_1z_2^2.
\end{eqnarray}
The PF operators are given by 
\begin{eqnarray}
\mathcal D_1&=&\theta_1^2-z_1(-2\theta_1+\theta_2)(-2\theta_1+\theta_2-1), \\
\no \mathcal D_2&=&(\theta_2-2\theta_1)\theta_2^2-z_2(-3\theta_2)(-3\theta_2-1)(-3\theta_2-2).
\end{eqnarray}

By using, once again, the PF system $\{\mathcal D_1,\theta_2\mathcal D_2\}$, we are able to find four point functions. Translating these to the $A$ model as in example 1, we arrive at 
\begin{eqnarray}
\lim_{t_2\rightarrow -\infty}C^{31}_{(4)}=-\frac{1}{3}\sum_{n\ge 1}e^{nt_1}.	
\end{eqnarray}	
We note that in this case, as above, the fiber curve has a triple intersection number $\langle C_2^3 \rangle=2$, so that we may define the prepotential, once again, by 
\begin{eqnarray}
	\frac{d^3 \mathcal F_{inst.}}{d t^3}= 
-3\lim_{t_2\rightarrow -\infty} C^{31}_{(4)}.
\end{eqnarray}
\end{example}
Hence, we have arrived at the expected instanton expansion for each of the two most trivial examples. We now turn to more general geometries.

\subsection{$K_S$ cases.}

We now demonstrate more fully the power of this approach by using the Calabi-Yau fourfold calculation to fully determine the prepotential on $K_{F_0}, K_{F_2}$ and $K_{dP_2}$ (up to polynomial terms of degree 2). In a previous work \cite{FJ}, the authors used a classical cohomology argument to produce extended Picard-Fuchs differential operators on $K_S$. These operators were then shown to reproduce the expected Yukawa couplings via the same techniques we used above on local $\p^1$. The disadvantage of the extended PF system is that there is not a simple closed form for the extended system on $K_S$. We will now show that through the fourfold formalism, all Yukawa couplings are produced automatically. We believe that this method should remain valid on every canonical bundle case. 

\begin{example}\normalfont
We begin with the canonical bundle over $F_0=\p^1\times \p^1$. The charge vectors for $X=K_{F_0}$ are 
\begin{eqnarray}
	\begin{pmatrix}
	l^1\\
	l^2
	\end{pmatrix}=
	\begin{pmatrix}
	-2&1&1&0&0\\
	-2&0&0&1&1
	\end{pmatrix}.
\end{eqnarray}
 The canonical bundle over the projective closure $\p(\mathcal{O}_{F_0}\oplus K_{F_0})=\bar X$ has the toric description
 \begin{eqnarray}
	\begin{pmatrix}
	0&-2&1&1&0&0&0\\
	0&-2&0&0&1&1&0\\
	-2&1&0&0&0&0&1
	\end{pmatrix}.
\end{eqnarray}
Let $Y$ be the mirror to $K_{\bar X}$. Then $Y$ is the family of hypersurfaces 
\begin{eqnarray}
	\{(u,v,y_4,y_5,y_6)\in \C^2\times (\C^*)^3:uv+1+z_2/(y_5y_6)+z_1z_2^2/(y_4 y_5^2y_6^2)+y_4+y_5+y_6=0\}.
\end{eqnarray}
As usual, there is a Picard-Fuchs system of differential operators whose solutions are the period integrals of $Y$:
\begin{eqnarray} \no
	\mathcal D_1&=& \theta_1^2-z_1(-2\theta_1-2\theta_2+\theta_3)(-2\theta_1-2\theta_2+\theta_3-1), \\ \no
	\mathcal D_2&=& \theta_2^2-z_2(-2\theta_1-2\theta_2+\theta_3)(-2\theta_1-2\theta_2+\theta_3-1), \\
	\mathcal D_3&=& \theta_3(\theta_3-2\theta_1-2\theta_2)-z_3(-2\theta_3)(-2\theta_3-1).
\end{eqnarray}
We let $t_1,t_2,t_3$ denote the logarithmic solutions.
As derived in section 3, we consider the extended Picard-Fuchs system $\{\mathcal D_1,\mathcal D_2,\theta_3\mathcal D_3\}$. Recall that the period integrals of $M$, the mirror of $\mathcal O (H_F)\oplus \mathcal O(-2H_F)\rightarrow \p (\mathcal O\oplus \mathcal O\oplus K_{F_0})$, coincide with the solutions of $\{\mathcal D_1,\mathcal D_2,\theta_3\mathcal D_3\}$. Set 
\begin{eqnarray}
Y^{mnp}_{(k)}&=&\int_M\Omega\wedge\nabla^m_{\delta_{z_1}}\nabla^n_{\delta_{z_2}}\nabla^p_{\delta_{z_3}} \Omega, \ m+n+p=k, \ k \in \{4,5\}, \\
\no  Y^{mnp}_{(k)}&=&0, \ k \le 3.
\end{eqnarray}
Using the procedure detailed above, we can fully determine all 14 of the $B$ model Yukawa couplings $Y^{mnp}_{(4)}$ on $M$. As in the local $\p^1$ case, we have to convert these couplings to the $A$ model (remembering that these functions transform as rank 4 tensors) and then take the limit $t_3\rightarrow -\infty$ in order to recover the correct Yukawa couplings. Let $C^{ijk}_{(4)}$ denote the $A$ model couplings on $K_{\bar X}$. We find, in particular, 
\begin{eqnarray} 
\label{F0fourpoint}
\no
	\lim_{t_3\rightarrow -\infty}C^{301}_{(4)}&=&-\frac{1}{2}\big(-2q_1-4q_1q_2-2q_1^2-48q_1^2q_2-6q_1q_2^2-2q_1^3-\dots\big), \\
	\lim_{t_3\rightarrow -\infty}C^{211}_{(4)}&=&-\frac{1}{2}\big(-4q_1q_2-24q_1^2q_2-12q_1q_2^2-\dots\big),
\end{eqnarray}
  where $q_i=e^{t_i}$. These, and the other two so-computed couplings, have exactly the instanton expansion expected, up to the scaling $-1/2$, which was derived in section 3.
This means that we should define a prepotential $\mathcal F$ for this space by the equations
  \begin{equation}
  \frac{\partial^3 \mathcal F}{\partial t_1^i \partial t_2^j}=-2 \lim_{t_3\rightarrow -\infty}C^{ij1}_{(4)}, \ i+j=3.
 \end{equation} 
 \end{example}

\begin{example} \normalfont
Next, we outline the construction for  $K_{F_2}$, the canonical bundle over the second Hirzebruch surface. The charge vectors for $X=K_{F_2}$ are 
\begin{eqnarray}
	\begin{pmatrix}
	l^1\\
	l^2
	\end{pmatrix}=
	\begin{pmatrix}
	-2&1&1&0&0\\
	0&0&-2&1&1
	\end{pmatrix}.
\end{eqnarray}
Charge vectors of $\p(\mathcal{O}_{F_2}\oplus K_{F_2})=\bar X$:
\begin{eqnarray}
	\begin{pmatrix}
	0&-2&1&1&0&0&0\\
	0&0&0&-2&1&1&0\\
	-2&1&0&0&0&0&1
	\end{pmatrix}.
\end{eqnarray}
The mirror $Y$ to $K_{\bar X}$ is 
\begin{eqnarray}
	\{(u,v,y_2,y_4,y_5)\in \C^2\times (\C^*)^3:uv+1+y_2+z_1y_2^2/y_4+y_4+y_5+z_2y_4^2/y_5+z_3/y_2=0\}.
\end{eqnarray}
The Picard-Fuchs system on $Y$ is in this case:
\begin{eqnarray} \no
	\mathcal D_1&=& \theta_1(\theta_1-\theta_2)-z_1(\theta_3-2\theta_1)(\theta_3-2\theta_1-1), \\ \no
	\mathcal D_2&=& \theta_2^2-z_2(\theta_1-2\theta_2)(\theta_1-2\theta_2-1), \\
	\mathcal D_3&=& \theta_3(\theta_3-2\theta_1)-z_3(-2\theta_3)(-2\theta_3-1).
\end{eqnarray}
Take $t_1,t_2,t_3$ to be the mirror map.
We work with the extended Picard-Fuchs systems $\{\mathcal D_1,\mathcal D_2,\theta_3\mathcal D_3\}$, whose solutions are the same as the period integrals of the mirror of $\mathcal O(H_F)\oplus \mathcal O(-2H_F)\rightarrow \p(\mathcal O\oplus \mathcal O \oplus K_{F_2})$.
We again compute the $B$ model Yukawa couplings $Y^{mnp}_{(4)}$ of $M$, and let $C^{ijk}_{(4)}$ be $A$ model couplings of $K_{\bar X}$. Then 
\begin{eqnarray} 
\label{F2fourpoint}
\no
	\lim_{t_3\rightarrow -\infty}C^{301}_{(4)}&=&-\frac{1}{2}\big(-2q_1-2q_1^2-2q_1q_2-32q_1^2q_2-2q_1^3-2q_1^4-\dots\big), \\ \no
	\lim_{t_3\rightarrow -\infty}C^{211}_{(4)}&=&-\frac{1}{2}\big(-2q_1q_2-16q_1^2q_2-54q_1^3q_2-2q_1^2q_2^2-\dots\big),\\ \no
	\lim_{t_3\rightarrow -\infty}C^{121}_{(4)}&=&-\frac{1}{2}\big(-2q_1q_2-8q_1^2q_2-2q_1^2q_2^2-18q_1^3q_2- \dots \big),\\ 
	\lim_{t_3\rightarrow -\infty}C^{031}_{(4)}&=&-\frac{1}{2}\big(-2q_1q_2-4q_1^2	q_2-6q_1^3q_2-2q_1^2q_2^2- \dots \big)
	\end{eqnarray}
We note, in particular, that the instanton number $N_{0,1}$ which was was computed to be $1/2$ in \cite{CKYZ} via localization, has a value of 0 in our calculation. This is in accordance with the localization computation performed in section 3.

Then the prepotential $\mathcal F_{inst.}$ is found by 
  \begin{equation}
  \frac{\partial^3 \mathcal F_{inst.}}{\partial t_1^i \partial t_2^j}=-2\lim_{t_3\rightarrow -\infty}C^{ij1}_{(4)}, \ i+j=3.
 \end{equation} 
 \end{example}

\begin{example}
\normalfont
Next, we briefly present the same computational procedure carried out on $K_{dP_2}$. Recall that this is defined by the vectors
\begin{eqnarray}
	\begin{pmatrix}
	-1&1&-1&1&0&0 \\
	-1&-1&1&0&0&1 \\
	-1&0&1&-1&1&0
	\end{pmatrix};
\end{eqnarray}
$K_{dP_2}$ is the canonical bundle over the blowup of $\p^2$ at two points.

Then we can immediately write the corresponding vectors for the 4fold over $K_{dP_2}$, namely $K_{\p(\mathcal O_{dP_2}\oplus K_{dP_2})}$:
\begin{eqnarray}
	\begin{pmatrix}
	0&-1&1&-1&1&0&0&0 \\
	0&-1&-1&1&0&0&1&0 \\
	0&-1&0&1&-1&1&0&0 \\
	-2&1&0&0&0&0&0&1
	\end{pmatrix}.
\end{eqnarray}
Let $Y$ be the mirror to this fourfold. $Y$ is given by
\begin{eqnarray}
Y=\{uv+	1+z_1y_4y_5/y_3+y_3+y_4+y_5+z_2y_3y_4/y_5+z_3/(y_3y_5)+z_4/y_4=0\}
\end{eqnarray}
The Picard-Fuchs system for period integrals on $Y$ consists of six order two operators
\begin{eqnarray}\no
\mathcal D_1&=&(\theta_1-\theta_2)(\theta_1-\theta_3)-z_1(-\theta_1-\theta_2-\theta_3+\theta_4)(-\theta_1+\theta_2+\theta_3), \\ \no
\mathcal D_2&=&(\theta_2-\theta_1+\theta_3)\theta_2-z_2(-\theta_1-\theta_2-\theta_3+\theta_4)(-\theta_2+\theta_1), \\ \no
\mathcal D_3&=&(\theta_3-\theta_1+\theta_2)\theta_3-z_3(-\theta_1-\theta_1-\theta_3+\theta_4)(-\theta_3+\theta_1), \\ \no
\mathcal D_4&=&(\theta_1-\theta_3)\theta_2-z_1z_2(-\theta_1-\theta_1-\theta_3+\theta_4)(-\theta_1-\theta_1-\theta_3+\theta_4-1), \\ \no
\mathcal D_5&=&(\theta_1-\theta_2)\theta_3-z_1z_3(-\theta_1-\theta_1-\theta_3+\theta_4)(-\theta_1-\theta_1-\theta_3+\theta_4-1), \\ \no
\mathcal D_6&=&(\theta_4-\theta_1-\theta_2-\theta_3)\theta_4-z_4(-2\theta_4)(-2\theta_4-1).
\end{eqnarray}
Let $t_1,\dots,t_4$ be the logarithmic solutions. 
We define the fourpoint functions $Y^{mnpq}_{(4)}$ in exact analogy with the earlier cases. We then solve for these fourpoint functions using the relations from the extended Picard-Fuchs system 
\begin{eqnarray}
	\{\mathcal D_1,\dots,\mathcal D_5, \theta_4\mathcal D_6\}.
\end{eqnarray}

 After transforming to the $A$ model (into functions $C^{mnpq}_{(4)}$)   and taking the large fiber limit, what we find is perfect agreement on all Yukawa couplings for the del Pezzo. We write here the first two such couplings:
\begin{eqnarray}\no
	\lim_{t_4\rightarrow -\infty}C^{3001}_{(4)}&=&-\frac{1}{2}\big(q_1+q_1^2-2q_1q_2-2q_1q_3+3q_1q_2q_3-2q_1^2q_2^2-2q_1^2q_3^2-32q_1^2q_2q_3+\dots\big) \\
	\lim_{t_4\rightarrow -\infty}C^{2101}_{(4)}&=&-\frac{1}{2}\big(-2q_1q_2+3q_1q_2q_3-16q_1^2q_2q_3-2q_1^2q_2^2+\dots \big)
\end{eqnarray}
\end{example}
These are as expected, up to the overall $-1/2$, which we predicted in Section 3. This  means that we recover the right instanton expansion via the normalization
\begin{eqnarray}
\frac{\partial^3 \mathcal F_{inst.}}{\partial t_1^i\partial t_2^j\partial t_3^k}=-2\lim_{t_4\rightarrow -\infty}C^{ijk1}_{(4)}, \ i+j+k=3
\end{eqnarray}

\section{Fourfold constructions for threefolds with  $b_4=0$.}

From the above, we have seen that while we can recover much additional information by using the projective closure plus canonical bundle technique, this seems to be unsuitable of there are too many noncompact divisors in the uncompactified geometry. The reason for this is as follows. If we attempt a straightforward projective closure procedure on a space with three or more noncompact divisors, the Poincare polynomial is badly behaved, and we are thus unable to use the technology introduced above in the computation of fourpoint functions. In particular, any local Calabi-Yau satisfying $\dim H_2(X,\Z)>1$, $\dim H_4(X,\Z)=0$ has at least three noncompact divisors, so we need new methods of analysis for such spaces.

With these difficulties in mind, we will develop tools tailor made to address this problem. In fact, we are able to show that for a large class of examples, by performing a partial compactification followed by a flop, we can reduce the problem to a $K_S$ type case. Then we have only to refer back to the methods introduced in the preceding sections on $K_S$, flop the resulting Yukawas back and take the appropriate limits to recover the Yukawa couplings on the geometry of interest. We will work through several examples to get a feel for the computational techniques.   

\subsection{The two one parameter cases.}

\begin{example}\normalfont We begin with the conifold, $X_1=\mathcal O(-1)\oplus \mathcal O(-1)\longrightarrow \p^1$. While the Yukawa coupling for the conifold has been derived above through simpler means, we present this example as a template for the types of methods we will use in the sequel.  

First, we reemphasize that the basic reason that local mirror symmetry (that is, local mirror symmetry via Picard-Fuchs systems) breaks down for the conifold is that there is no 4 cycle on this space. Hence, the PF system on the mirror cannot have a double log solution, and therefore we cannot recover an instanton expansion. 

With this as motivation, we will consider a simple noncompact threefold which contains the conifold geometry, as well as a new four cycle. The candidate `compactification', which we call $\tilde X_1$, is depicted in Figure \ref{p1completion2}, and is defined by the toric charge vectors
\begin{eqnarray}
	\begin{pmatrix}
	l^1 \\
	l^2
	\end{pmatrix}=
	\begin{pmatrix}
	1&1&-1&-1&0 \\
	-3&0&1&1&1
	\end{pmatrix}.
\end{eqnarray}
\begin{figure}[t]
\centering
\input{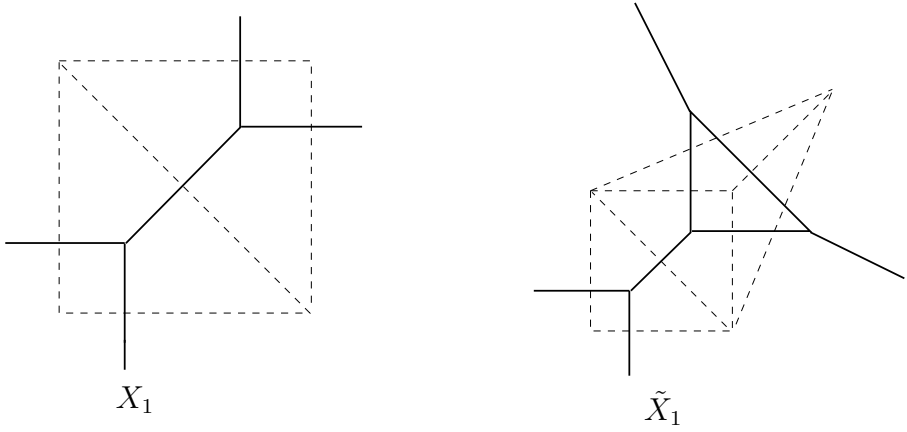}
\caption{Toric diagram for the addition of a 4 cycle to $\mathcal O(-1)\oplus \mathcal O(-1)\longrightarrow \p^1$.}
\label{p1completion2}
\end{figure}
Now, we want to connect this to our previous constructions, i.e. the canonical bundle over a surface case. But this is easy, because $\tilde X_1$ admits a flop to $K_{F_1}$:
\begin{eqnarray}
	\begin{pmatrix}
	-l^1 \\
	l^1+l^2
	\end{pmatrix}=
	\begin{pmatrix}
	-1&-1&1&1&0 \\
	-2&1&0&0&1
	\end{pmatrix}.
\end{eqnarray}
That is, $\tilde X_1^{flop}\cong K_{F_1}$. 
Now we use the machinery of previous sections. Let $K_{\bar K_{F_1}}$ be the noncompact fourfold associated to $K_{F_1}$, defined by charge vectors
\begin{eqnarray}
	\begin{pmatrix}
	0&-1&-1&1&1&0&0 \\
	0&-2&1&0&0&1&0\\
	-2&1&0&0&0&0&1
	\end{pmatrix}.
\end{eqnarray}
 We denote the mirror of $K_{\bar{K}_{F_1}}$ by $Y$. This procedure is summarized by the following sequence of operations:
\begin{displaymath}
\xymatrix{X_1\ar[r] & \tilde{X_1}\ar[r] & \tilde{X}_1^{flop}\cong K_{F_1}\ar[r]& K_{\bar K_{F_1}} }
\end{displaymath}
Then exactly as in the $K_{F_0}$ case, we can compute $B$ model fourpoint functions $Y^{mnp}_{(4)}(z_1,z_2,z_3)$ on $Y$ by using the Picard-Fuchs system. Here $z_1,z_2,z_3$ are the local variables on the complex structure moduli space of $Y$.

The next step is to carry the $B$ model fourpoint functions across the flop on the $B$ model, defined by the change of variables
\begin{equation}
z_1=w_1^{-1}, \ z_2=w_1w_2, \ z_3=w_3.
\end{equation} 
We let $\tilde Y$ be the manifold we get by using the flop transformation on $Y$. Here, we have to remember that the $Y^{mnp}_{(4)}(z_1,z_2,z_3)$ transform as rank 4 tensors. Then, we have fourpoint functions $\tilde{Y}^{mnp}_{(4)}(w_1,w_2,w_3)$  on $\tilde Y$, which is also the mirror of the fourfold over $\tilde{X}_1$. Let $t_1,t_2,t_3$ be the logarithmic solutions of the Picard-Fuchs system on $\tilde Y$. 

Next, use the inverse mirror map $w(t)$ to convert the $\tilde{Y}^{mnp}_{(4)}(w_1,w_2,w_3)$ into $A$ model fourpoint functions $\tilde{C}^{mnp}_{(4)}$ on the fourfold over $\tilde{X_1}$, again taking the tensor property into account. And then finally, we can recover the expected Yukawa threepoint function $C_{X_1}$ on $X_1$ in the limit as $t_2,t_3\rightarrow -\infty$. 

Since this whole procedure has been rather complicated, we summarize the various steps in the following diagram.
\begin{displaymath}
\xymatrix{ Y^{mnp}_{(4)} \ar[r] & \tilde{Y}^{mnp}_{(4)} \ar[r] & \tilde{C}^{mnp}_{(4)} \ar[r] & C_{X_1}  \\ 
     Y \ar[r] &   \tilde Y \ar[r]    &  K_{\bar{K}_{\tilde{X}_1}} \ar[r]  &  X_1  }
\end{displaymath}
The functions along the top line are the fourpoint functions of the corresponding spaces on the bottom line.
On the bottom line, the first arrow is given by the flop, the second by the mirror map, and the third by taking the double limit $t_2,t_3\longrightarrow -\infty$. These two limits are to be understood as first taking the size of the $\p^2\hookrightarrow \tilde{X_1}$ to infinity, and then taking the limit of the large compactification fiber (that is, the limit in which the noncompact fourfold becomes a noncompact threefold). The result of this is  
\begin{eqnarray}
	C_{X_1}=\lim_{t_2,t_3\rightarrow -\infty}\tilde{C}^{301}_{(4)}=-\frac{1}{2}\frac{e^{t_1}}{1-e^{t_1}}.
\end{eqnarray}
Here, an extra factor $-\frac{1}{2}$ appear  because we have used a
$\mathbb{P}^{1}$ compactification.
\end{example}

\begin{example}
\label{zerominustwo}
\normalfont
For our next example, we revisit $X_2=\mathcal O\oplus \mathcal O (-2)\longrightarrow \p^1$. Once again, though we have already worked out the Yukawa coupling for this case through the fourfold, we now want to take a look at another way of deriving this fourpoint function. The reason is that this new viewpoint is the one that will prove to be naturally applicable to the general case.

As in the previous example, we want to add a four cycle at some convenient location in the geometry in order to recover the instanton expansion. In contrast with the previous example, we also have to simultaneously compactify the one parameter noncompact deformation space of this $\p^1$. The only choice that satisfies both of these criteria is $K_{F_0}$:
\begin{eqnarray}
	\begin{pmatrix}
	-2&1&1&0&0& \\
	-2&0&0&1&1&
	\end{pmatrix}.
\end{eqnarray}
This is depicted in Figure \ref{p1completionsecond}.

Now, we have already done the fourfold calculation on $K_{F_0}$, so we only have to refer to the Yukawa couplings above, Eqn.(\ref{F0fourpoint}). Let $C^{(301)}_{(4)}$ be that taken from Eqn.(\ref{F0fourpoint}), and let $t_1,t_2$ be the sizes of the two $\p^1$s in $F_0$. Then we find, in the relevant limit,
\begin{eqnarray}
\lim_{t_2,t_3\rightarrow -\infty}C^{(301)}_{(4)}=-\frac{1}{2}\big(-2q_1-2q_1^2-2q_1^3-\dots\big)
\end{eqnarray}
\end{example}
So, even after taking into account the extra factor $-1/2$ from the fourfold compactification, we still see that the instanton expansion is twice what we expect it should be. The reason for this is, however, easy to see. In the $K_{F_0}$ geometry, if we perform a direct localization calculation, then we find that there are two curves in each of the two homology classes, which is obvious from the toric diagram. Thus, in order to recover the correct expansion, we have to remove by hand the excess state. After doing this we indeed get what we were expecting, complete with the overall negative sign \cite{I}.

\subsection{Higher parameter examples.}

We now present our computational scheme in its general form. The basic idea is to complete all curves with normal bundle $\mathcal O\oplus \mathcal O(-2)$ by using the $F_0$-type compactification given in Example \ref{zerominustwo}. This kind of example was first considered in \cite{DFG}. After doing this, we find that we can recover all Yukawa couplings using the same trick as above, i.e. by flopping to a canonical bundle case and then taking the noncompact fourfold over the canonical bundle. This method works well for a reasonably broad class of geometries. We will carefully go through the details of two more examples.
\begin{example}
\label{example7}
\normalfont
We take $X$ to be a local threefold with $\dim H_2(X,\Z)=2,\dim H_4(X,\Z)=0$ defined by the charge vectors 
\begin{eqnarray}
	\begin{pmatrix}
	l^1  \\
	l^2
	\end{pmatrix}=
	\begin{pmatrix}
	-2&1&1&0&0 \\
	1&-1&0&1&-1
	\end{pmatrix}.
\end{eqnarray}
There are two curves $C_t$,$C_s$ corresponding to the vectors $l^1,l^2$ respectively. From the vectors we can read off that $\mathcal N_{C_t/X}\cong \mathcal O \oplus \mathcal O(-2)$,$\mathcal N_{C_s/X}\cong \mathcal O(-1) \oplus \mathcal O(-1)$. There is one more curve $C_{s+t}$ which also has a normal bundle of $\mathcal O(-1) \oplus \mathcal O(-1)$.

Then, from the examples of the previous section it is clear that we only need to compactify the $C_t$ curve in order to derive a complete set of Yukawa couplings using the fourfold construction. Let the space we get by compactifying the $C_t$ family be denoted by $\tilde{X}$. Then $\tilde{X}$ is given by the charge vectors
\begin{eqnarray}
	\begin{pmatrix}
	l^0 \\
	l^1  \\
	l^2
	\end{pmatrix}=
	\begin{pmatrix}
  -2&0&0&0&1&1\\
	-2&1&1&0&0&0 \\
	1&-1&0&1&-1&0
	\end{pmatrix}.
\end{eqnarray}
In order to convert this to a canonical bundle case, we can flop to $K_{dP_2}$:
\begin{eqnarray}
	\begin{pmatrix}
	l^0+l^2 \\
	l^1+l^2  \\
	-l^2
	\end{pmatrix}=
	\begin{pmatrix}
  -1&-1&0&1&0&1\\
	-1&0&1&1&-1&0 \\
	-1&1&0&-1&1&0
	\end{pmatrix}.
\end{eqnarray}
This is depicted in Figure \ref{dpfour}. 
\begin{figure}[t]
\centering
\input{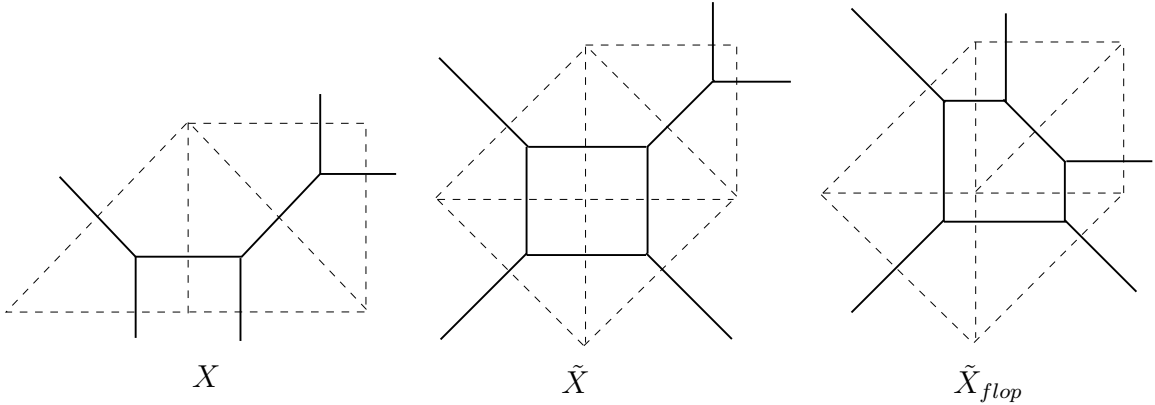}
\caption{Geometric transformations of $X$.}
\label{dpfour}
\end{figure}
Now that we have a canonical bundle case, we can proceed as usual with the fourfold calculation. Let $t_3$ be the K\"ahler parameter corresponding to $\p(\mathcal O_{dP_2} \oplus K_{dP_2})$. Since we have already worked out all the fourpoint functions for the del Pezzo, we can just use these and flop them back to find the appropriate Yukawa couplings for the present case. After doing this, we find the threepoint Yukawa couplings on the original geometry in the large fiber limit:
\begin{eqnarray}
	\lim_{t_0,t_3\rightarrow -\infty}C^{301}_{(4)}&=&-\frac{1}{2}\Big(-2\frac{e^{t_1}}{1-e^{t_1}}+\frac{e^{t_1+t_2}}{1-e^{t_1+t_2}}\Big), \\
		\lim_{t_0,t_3\rightarrow -\infty}C^{210}_{(4)}&=&-\frac{1}{2}\frac{e^{t_1+t_2}}{1-e^{t_1+t_2}}, \\
			\lim_{t_0,t_3\rightarrow -\infty}C^{121}_{(4)}&=&-
\frac{1}{2}\frac{e^{t_1+t_2}}{1-e^{t_1+t_2}}, \\
				\lim_{t_0,t_3\rightarrow -\infty}C^{031}_{(4)}&=&-\frac{1}{2}\Big(1+\frac{e^{t_2}}{1-e^{t_2}}+\frac{e^{t_1+t_2}}{1-e^{t_1+t_2}}\Big).
\end{eqnarray}
We again see the same phenomena from the earlier examples. First, the overall $-1/2$ comes from $\mathbb{P}^{1}$ compactification associated to 
$K_{dP_2}$. Secondly, we have to remove by hand the overcounted state which is represented by 
\begin{eqnarray}
	-2\frac{e^{t_1}}{1-e^{t_1}}.
\end{eqnarray}
  After this, we find complete agreement with the expected instanton information on this space \cite{DFG}. In other words, we may define a prepotential for this example by
\begin{eqnarray}
	\frac{\partial^3 \mathcal F_{inst.}}{\partial t_1^i\partial t_2^j}=
-2\lim_{t_0,t_3\rightarrow -\infty}C^{ij1}_{(4)}, \ \ i+j=3
\end{eqnarray}
up to the overcounted $(-2,0)$ curve.
\end{example}

\begin{example}\normalfont
Finally, we want to consider a rather complicated example, which will help to illustrate the general procedure. The space we have in mind was considered in \cite{dFS}, and is specified by charge vectors 
\begin{eqnarray}
\begin{pmatrix}
l^1 \\ l^2 \\ l^3
\end{pmatrix}=
\begin{pmatrix}
1&0&0&1&-1&-1 \\
0&1&0&-1&1&-1 \\
0&0&1&-1&-1&1 
\end{pmatrix}.
\end{eqnarray}
We denote this by $X$. Note that $\dim H_2(X)=3, \dim H_4(x)=0$, and the three curves in $X$ have a single point of intersection. The toric graph of this space, complete with triangulation, is shown in Figure \ref{trivalent1}.
\begin{figure}[t]
\centering
\input{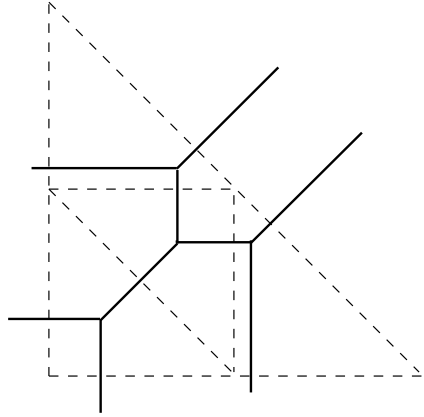}
\caption{The trivalent curve problem.}
\label{trivalent1}
\end{figure}

Now, from the previous examples of this type, the general idea we have followed is to `compactify' the curves with normal bundle $\mathcal O \oplus \mathcal O(-2)$ via the scheme we originally used for the one parameter space $\mathcal O \oplus \mathcal O(-2)\rightarrow \p^1$. In our present situation, such curves are not evident, but we can make them manifest by performing a flop transition. We call the resulting space $X_{flop}$, and its charge vectors are 
\begin{eqnarray}
\begin{pmatrix}
-l^1 \\ l^2+l^1 \\ l^3+l^1
\end{pmatrix}=
\begin{pmatrix}
-1&0&0&-1&1&1 \\
1&1&0&0&0&-2 \\
1&0&1&0&-2&0 
\end{pmatrix}.
\end{eqnarray}
Then, we see that the second and third curves have normal bundle $\mathcal O \oplus \mathcal O(-2)$. 

In order to keep the calculation from getting too unwieldy, we will only compactify one of the $(-2,0)$ curves, and proceed with the calculation on the resulting space. Upon doing this, we get a new space $\tilde X_{flop}$ specified by charge vectors 
\begin{eqnarray}
\begin{pmatrix}
k^0 \\ k^1 \\ k^2 \\ k^3 
\end{pmatrix}=
\begin{pmatrix}
0&0&0&1&-2&0&1 \\
-1&0&0&-1&1&1&0 \\
1&1&0&0&0&-2&0 \\
1&0&1&0&-2&0&0 
\end{pmatrix}.
\end{eqnarray}
From the charge vectors alone, it is a bit hard to see what is going on, so we have given a diagrammatic representation of this procedure in Figure
\ref{trivalent2}.
\begin{figure}[t]
\centering
\input{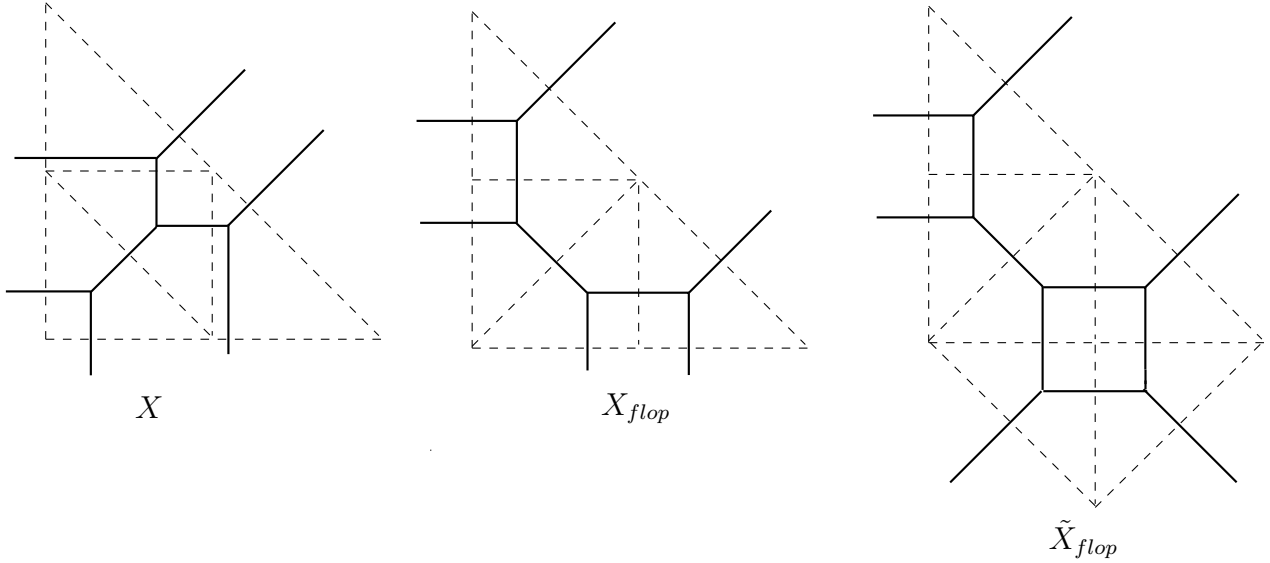}
\caption{Geometric manipulations of the trivalent curve.}
\label{trivalent2}
\end{figure}

Before diving into the details, let us briefly consider what exactly it is that we are expecting to learn through the study of this space $\tilde{X}_{flop}$. The only real difference between $\tilde X$ and $\tilde{X}_{flop}$ is that on the latter, a single $(-2,0)$ curve family has been compactified. If we look back at the original geometry $X$, this corresponds to `filling in' the curve information corresponding to the $l^1+l^3$ curve on $X$. Therefore, the predicted result is that the instanton expansion we find will enumerate curve data corresponding to the curves $l^1,l^2,l^3,l^1+l^2+l^3$, and $l^1+l^3$. That is, we will obtain all information corresponding to curves with normal bundle $\mathcal O(-1)\oplus \mathcal O(-1)$, plus the curve $l^1+l^3$ (which has normal bundle $\mathcal O\oplus \mathcal O(-2)$) that we have completed by using our compactification. Finally, this last curve should be counted with an overall $-2$ in the instanton sum, as a result of the type of compactification we are using. 

With that being said, let's proceed with the computation. The first thing we have to do is associate a noncompact fourfold to the above geometry $\tilde{X}_{flop}$. In the previous examples, we have done this by first reducing to a canonical bundle case and then compactifying the canonical bundle. While this can indeed be done here, we claim in the present case that it suffices to compactify the variable corresponding to the compact divisor in the geometry. From Figure \ref{trivalent2}, it is clear that there is exactly one compact divisor, namely $\p^1\times \p^1$, and moreover this corresponds to the fifth column of the matrix of charge vectors defining $\tilde{X}_{flop}$. One can see this by recalling the charge vectors for $K_{\p^1\times \p^1}$:
\begin{eqnarray}
	\begin{pmatrix}
	-2&1&1&0&0 \\
	-2&0&0&1&1
	\end{pmatrix}
\end{eqnarray}
Here, the divisor corresponding to the first column represents the $\p^1\times \p^1$, and we note that the fifth column of the charge vectors for $\tilde{X}_{flop}$ also contains two $-2$ entries.

Then, it is straightforward to write down the charge matrix of the fourfold over $\tilde{X}_{flop}$:   
\begin{eqnarray}
\begin{pmatrix}
m^0 \\ m^1 \\ m^2 \\ m^3 \\ m^4
\end{pmatrix}=
\begin{pmatrix}
0&0&0&0&1&-2&0&1&0 \\
0&-1&0&0&-1&1&1&0&0 \\
0&1&1&0&0&0&-2&0&0 \\
0&1&0&1&0&-2&0&0&0 \\
-2&0&0&0&0&1&0&0&1
\end{pmatrix}.
\end{eqnarray}
We denote the above space by $\hat X$. Let us consider a bit further why it is that we expect this fourfold to reproduce the instanton information we are looking for. Previously, most of our successful calculations have been done on a canonical bundle example. Notice first of all that we can perform two flops on $\tilde{X}_{flop}$ to reach a canonical bundle case: 
\begin{eqnarray}
	\begin{pmatrix}
	-1&0&0&0&-1&1&1 \\
	1&1&0&0&0&-2&0 \\
	0&-1&0&1&-1&1&0 \\
	0&1&1&-2&0&0&0
	\end{pmatrix}
\end{eqnarray}
This is done by first flopping the second vector of $\tilde{X}_{flop}$, and then flopping the fourth vector of the resulting space. On this matrix, it is clear that the fifth column corresponds to the single compact divisor, and furthermore all entries of the fifth column are less than or equal to 0, so that this is a $K_S$ case. This can also be seen by constructing the vertices for this manifold. Now, since the compactification variable is fixed across the flop, it should be sufficient to just work directly on the space $\hat X$ above. And indeed, this will turn out to be the case. 

Let $\hat Y$ be the mirror manifold to $\hat X$.
We omit the details, but merely note that there are 10 order two Picard-Fuchs operators $\{\mathcal D_1,\dots \mathcal D_{10}\}$ whose solution space describes the period integrals of $\hat Y$. Let $t_0,\dots,t_4$ be the logarithmic solutions of this system. As in all previous cases, we use the extended set of differential operators $\{\mathcal D_1,\dots \theta_5 \mathcal D_{10}\}$ in order to solve for the fourpoint functions of $\hat Y$. Let $\hat{Y}^{mnpqr}_{(4)}$ be the fourpoint functions so obtained. Then we first use the inverse of the mirror map $t_0,\dots,t_4$ to transform these fourpoint functions on $\hat Y$ into fourpoint functions $\hat{C}^{mnpqr}_{(4)}$ on $\hat X$. Next, we recover the threepoint functions $C^{npq}_{flop}$ on $X_{flop}$ in the double scaling limit:
\begin{eqnarray}
 C^{npq}_{flop}=\lim_{t_0,t_4\rightarrow -\infty}\hat{C}^{0,n,p,q,1}_{(4)}.	
\end{eqnarray}
 And lastly, we can compute the threepoint functions we are looking for, $C^{npq}$ on $X$, by reversing the flop transition on $ C^{npq}_{flop}$ (this function transforms as a rank 3 tensor). After all is said and done, we arrive at the threepoint functions for $X$. For brevity, we list only a representative subset of the results here:
\begin{eqnarray}
	C^{300}&=&-\frac{1}{2}\Big(\frac{e^{t_1}}{1-e^{t_1}}-2\frac{e^{t_1+t_3}}{1-e^{t_1+t_3}}+\frac{e^{t_1+t_2+t_3}}{1-e^{t_1+t_2+t_3}}\Big), \\
C^{030}&=&-\frac{1}{2}\Big(\frac{e^{t_2}}{1-e^{t_2}}+\frac{e^{t_1+t_2+t_3}}{1-e^{t_1+t_2+t_3}}\Big), \\
C^{201}&=&-\frac{1}{2}\Big(-2\frac{e^{t_1+t_3}}{1-e^{t_1+t_3}}+\frac{e^{t_1+t_2+t_3}}{1-e^{t_1+t_2+t_3}}\Big),	\\
C^{021}&=&-\frac{1}{2}\frac{e^{t_1+t_2+t_3}}{1-e^{t_1+t_2+t_3}}.		
\end{eqnarray}
From these functions, we can see many of the previously advertised features of the compactification scheme we have chosen. As expected, the $t_1+t_3$ curve appears with an overall $-2$ factor, from the $\p^1\times \p^1$ type compactification. Besides this, the expansion is missing both of the other double curve classes $t_1+t_2$ and $t_2+t_3$. In other words, for example, we would expect to find the term
\begin{eqnarray}
\label{wasuremono}
	-\frac{e^{t_2+t_3}}{1-e^{t_2+t_3}}
\end{eqnarray}
 in the expansion for $C^{021}$, since the $t_2+t_3$ curve has normal bundle $\mathcal O \oplus \mathcal O(-2)$. This can be seen from the topological vertex calculation \cite{I}. 
  
  Nonetheless, since the original space $X$ is pairwise symmetric under the exchange of any two of the curves with normal bundle $\mathcal O(-1)\oplus \mathcal O(-1)$, it is clear that we could have compactified either of the other two $(-2,0)$ curves and picked up the missing terms ala eqn.(\ref{wasuremono}). Therefore, up to the overall fraction $1/6$, we have arrived at the expected instanton expansion. 
  
The extra factor $-1/2$ appears in  the same way as in all previous examples.
    
\end{example}

\begin{figure}[t]
\centering
\input{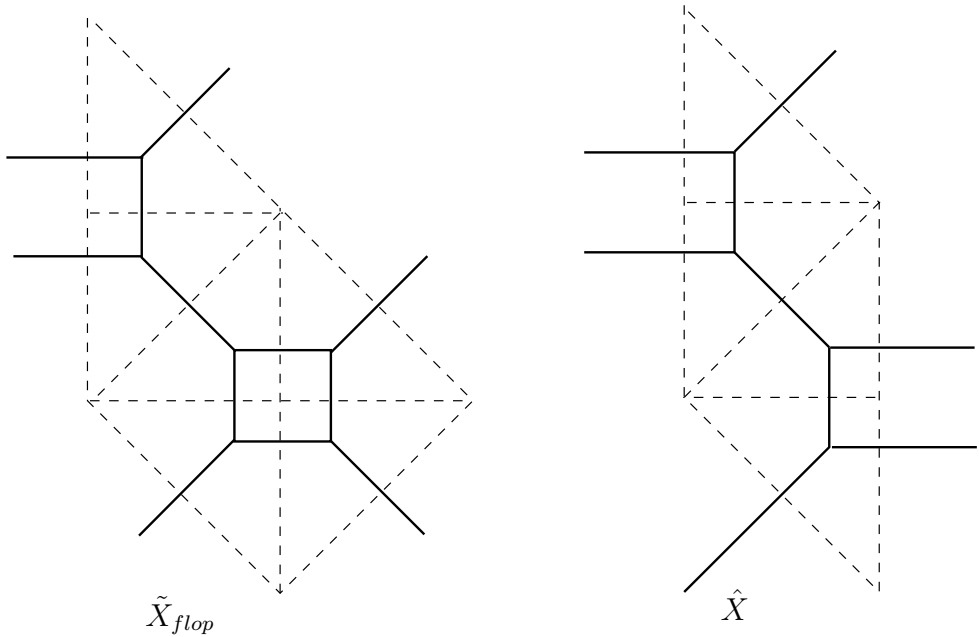}
\caption{Taking the other limit on $\tilde{X}_{flop}$.}
\label{otherlimit}
\end{figure}

Finally, to close this example, note that there is in fact more we can do with the space $\tilde{X}_{flop}$. That is, instead of taking the limit $t_0\rightarrow -\infty$, we can also consider the limit $t_3\rightarrow -\infty$. The result of this is shown in Figure \ref{otherlimit}.
This is a different Calabi-Yau, which we denote by $\hat X$, and we can use the same results above in this new limit in order to compute the threepoint functions on $\hat X$. Without going into any detail, we merely list two of the threepoint functions obtained this way: 
\begin{eqnarray}
	C_{\hat X}^{003}&=&-\frac{1}{2}\Big(-2\frac{e^{t_2}}{1-e^{t_2}}+\frac{e^{t_1+t_2}}{1-e^{t_1+t_2}}+\frac{e^{t_0+t_1+t_2}}{1-e^{t_0+t_1+t_2}}\Big), \\
	C_{\hat X}^{300}&=&-\frac{1}{2}\Big(\frac{e^{t_0+t_1}}{1-e^{t_0+t_1}}+\frac{e^{t_0+t_1+t_2}}{1-e^{t_0+t_1+t_2}}\Big).
\end{eqnarray}
In other words, the curve information corresponding to the term 
\begin{eqnarray}
	\frac{e^{t_0}}{1-e^{t_0}}
\end{eqnarray}
 is missing. This is expected, because we did not compactify this curve family. 
 
 \subsection{A word about sign conventions.}
 
 In the two preceding examples, the type of compactification we used was chosen according to topological vertex calculations \cite{I}. In the following, we present an argument that in some sense, the sign choice coming from the vertex computation is artificial (that is, it is extrinsic to the geometry).
 
Consider again Example \ref{example7}. The compactification used there, which was originally suggested in \cite{DFG}, was made so that we would find the following result for the instanton part of the prepotential:
\begin{eqnarray}
\label{instpart}
	\mathcal F^{inst}=\sum_{n>0}\frac{-e^{nt_1}+e^{n(t_1+t_2)}+e^{nt_2}}{n^3}
\end{eqnarray}
 Recall that $t_1$ was the complexified K\"ahler parameter for the curve with normal bundle $\mathcal O\oplus \mathcal O(-2)$, and the K\"ahler parameters $t_1+t_2$ and $t_2$ both correspond to curves with normal bundle $\mathcal O(-1)\oplus \mathcal O(-1)$. In other words, we have associated a minus sign to $(0,-2)$ curves, and a plus sign for $(-1,-1)$ curves. 
 
 However, we claim that from the geometry of Example \ref{example7} alone, this sign choice is not unique. For example, were we to use instead the instanton part  \begin{eqnarray}
\label{instpart2}
	\big(\mathcal F^{inst}\big)'=\sum_{n>0}\frac{e^{nt_1}+e^{n(t_1+t_2)}+e^{nt_2}}{n^3},
\end{eqnarray}
the answer would be equally `acceptable', in the following sense. We recall from \cite{JN},\cite{FJ} the conjecture that the $B$ model Yukawa couplings should be simple rational functions, such that the denominator consists of the components of the discriminant locus. Then, if we use either $\mathcal F^{inst}$ or $\big(\mathcal F^{inst}\big)'$ (together with the triple intersection numbers conjectured in \cite{FJ}), we find rational $B$ model Yukawa couplings of exactly the same level of complexity. Moreover, the resulting extended Picard-Fuchs system \cite{FJ} is also of roughly the same form. 

We will add further evidence to this claim in the appendix, where we construct the extended Picard-Fuchs system for the mirror of the trivalent curve for both choices of sign convention. Indeed, it turns out that in both cases, we find a system of nearly identical complexity.

\section{Conclusion.}
The main features of this paper are summarized as follows.

First, in \cite{FJ}, we  made use of the instanton expansion for $K_S$ cases in order to compute the allowed values for the classical triple intersection numbers; in the present work, through the use of the canonical bundle formula, we have carried out the computation of these numbers in a way that is more intrinsic to the geometry. 

Secondly, we have seen, besides the construction of the prepotential, the resolution of another problem encountered in \cite{FJ}. In \cite{FJ}, in order to construct the extended Picard-Fuchs system for $X$ such that $b_4(X)=0$, we took for granted the known instanton expansion from the topological vertex. Above, we have overcome this through the use of a special compactification scheme \cite{DFG} which is known to agree with the vertex result; the advantage of this is that, in principle, it applies to any $X$ with $b_4(X)=0$. 

We briefly mention some directions for future study. We are currently working to extend our results to non-nef toric varieties and their canonical bundles, e.g. $K_{F_n}$ for $n \ge 3$ and $\p(\mathcal O\oplus \mathcal O(k)\oplus \mathcal O(-2-k))$ for $k \ge 1$. In both cases, we will need to take advantage of the machinery of generalized mirror symmetry (ala Jinzenji, Iritani, Coates-Givental) in order to complete the calculation. We hope to report on these matters in future work.

\section*{Appendix A: Extended Picard-Fuchs System of Trivalent Toric Graph}
Since this model is symmetric under the permutation of the three K\"ahler 
parameters, we present a minimal set of formulas for brevity.
First, we start from the A-model Yukawa couplings obtained from the body 
of this paper with the constant term predicted in \cite{FJ}: 
\begin{eqnarray}
Y_{111} &=& \frac{e^{t^{1}}}{1-e^{t^{1}}}- 
\frac{e^{t^{1}+t^{2}}}{1-e^{t^{1}+t^{2}}}-
\frac{e^{t^{1}+t^{3}}}{1-e^{t^{1}+t^{3}}}+
\frac{e^{t^{1}+t^{2}+t^{3}}}{1-e^{t^{1}+t^{2}+t^{3}}},\nonumber\\
Y_{112} &=& -\frac{e^{t^{1}+t^{2}}}{1-e^{t^{1}+t^{2}}}+
\frac{e^{t^{1}+t^{2}+t^{3}}}{1-e^{t^{1}+t^{2}+t^{3}}},\nonumber\\
Y_{123} &=&\frac{1}{2}+\frac{e^{t^{1}+t^{2}+t^{3}}}{1-e^{t^{1}+t^{2}+t^{3}}}.
\label{A}
\end{eqnarray}
We can also read off the mirror maps from the ordinary Picard-Fuchs 
system of trivalent toric graph as follows:
\begin{equation}
t_{1}:=\log(z_{1})+\log(\frac{1}{2}(1+\sqrt{1-4z_{2}z_{3}}))-\log(\frac{1}{2}(1+\sqrt{1-4z_{1}z_{2}}))-\log(\frac{1}{2}(1+\sqrt{1-4z_{1}z_{3}})).
\label{m}
\end{equation}
Then, we can obtain 3-fold version of the A-model Gauss-Manin system for
 this toric-graph, as was defined in \cite{FJ}. After transforming this 
Gauss-Manin system into the B-model by the above mirror maps, we can obtain 
the extended Picard-Fuchs system $\{{\cal D}_{1}, {\cal D}_{2}, {\cal D}_{3}\}$
as relations of the B-model Gauss-Manin system. 
Here, we present ${\cal D}_{1}$ as follows. ${\cal D}_{2}, {\cal D}_{3}$ 
are obtained from the cyclic permutation of the subscripts $1,\;2,\;3$ of
${\cal D}_{1}$:  
\begin{eqnarray}
&&{\cal D}_{1}:=
(-5z_1^2z_3^2+2z_1^3z_2^2+5z_1z_2-10z_1^2z_3z_2+4z_1^3z_2^2z_3+2z_2z_3^2z_1+2z_1^3z_3^2+z_1+\nonumber\\&& 8z_1^3z_3z_2+
6z_1^2z_2z_3^2+6z_1^2z_2^2z_3-6z_1z_2z_3-8z_1^2z_2^2z_3^2-1+2z_2^2z_3z_1-5z_1^2z_2^2-4z_1^2z_3+\nonumber\\&&4z_1^3z_2z_3^2+
5z_1z_3-4z_1^2z_2)\theta_1^2+
((-4z_1z_3+2z_1z_2+2z_1^2z_3-4z_1^3z_3^2-4z_2z_3^2z_1-\nonumber\\&&4z_2^2z_3z_1-2z_1^2z_2-12z_1^2z_2z_3^2-4z_1^2z_2^2z_3+16z_1^2z_2^2z_3^2+4z_1^3z_2^2+10z_1^2z_3^2-6z_1^2z_2^2-\nonumber\\&&
8z_1^3z_2z_3^2+8z_1^3z_2^2z_3+8z_1z_2z_3)\theta_2+
z_1z_3+4z_1^3z_2^2z_3-2z_1^2z_2^2z_3+z_1z_2+\nonumber\\&&(-4z_1z_2+2z_1z_3-2z_1^2z_3+4z_1^3z_3^2-
4z_2z_3^2z_1-4z_2^2z_3z_1+2z_1^2z_2-4z_1^2z_2z_3^2-12z_1^2z_2^2z_3+\nonumber\\&&16z_1^2z_2^2z_3^2-4z_1^3z_2^2-6z_1^2z_3^2+
10z_1^2z_2^2+8z_1^3z_2z_3^2-8z_1^3z_2^2z_3+8z_1z_2z_3)\theta_3-
2z_2z_3^2z_1-\nonumber\\&&z_1^2z_3-2z_2^2z_3z_1-
3z_1^2z_2^2-z_1^2z_2-2z_1^2z_3z_2+2z_1z_2z_3+2z_1^3z_3^2+2z_1^3z_2^2-
3z_1^2z_3^2+8z_1^2z_2^2z_3^2+\nonumber\\&&4z_1^3z_2z_3^2-2z_1^2z_2z_3^2)\theta_1+
(4z_1^3z_2z_3^2-4z_1^3z_2^2z_3-2z_1z_2z_3-z_1z_2+2z_1^2z_2z_3^2-8z_1^2z_2^2z_3^2+\nonumber\\&&2z_1^2z_2^2z_3+2z_2z_3^2z_1+2z_2^2z_3z_1-
z_1^2z_3-2z_1^3z_2^2-3z_1^2z_3^2+3z_1^2z_2^2+z_1^2z_2+2z_1^3z_3^2+\nonumber\\&&z_1z_3)\theta_3+ 
(-8z_1^2z_2^2z_3^2+2z_2z_3^2z_1+6z_1^2z_2z_3^2+4z_1^3z_2^2z_3+
2z_1^3z_2^2-2z_1z_2z_3+4z_1^3z_2z_3^2+\nonumber\\&&2z_1^2z_3-5z_1^2z_3^2+2z_1^2z_3z_2-2z_1^2z_2^2z_3+2z_1^3z_3^2+2z_2^2z_3z_1-8z_1^3z_3z_2-
z_1^2z_2^2)\nonumber\\&&
(-4z_1^3z_2z_3^2+4z_1^3z_2^2z_3-2z_1z_2z_3+
2z_2z_3^2z_1+2z_1^2z_2z_3^2-8z_1^2z_2^2z_3^2+2z_1^2z_2^2z_3+3z_1^2z_3^2+\nonumber\\&&2z_2^2z_3z_1+z_1^2z_3+2z_1^3z_2^2-2z_1^3z_3^2+
(4z_2z_3^2z_1-4z_1^2z_2-2z_1z_2+4z_1^2z_2z_3^2-4z_1^3z_3^2-4z_1^2z_3-\nonumber\\&&16z_1^2z_2^2z_3^2-4z_1^3z_2^2+4z_2^2z_3z_1+
16z_1^3z_3z_2+6z_1^2z_3^2+6z_1^2z_2^2-2z_1z_3+4z_1^2z_3z_2-8z_1^3z_2z_3^2-\nonumber\\&&8z_1^3z_2^2z_3+
4z_1^2z_2^2z_3-4z_1z_2z_3+2z_1)\theta_3-
3z_1^2z_2^2-z_1^2z_2+z_1z_2-z_1z_3)\theta_2+\nonumber\\&&
(-5z_1^2z_2^2+2z_2^2z_3z_1+2z_1^3z_3^2+2z_1^2z_2-8z_1^2z_2^2z_3^2+2z_1^2z_3z_2+2z_1^3z_2^2-2z_1z_2z_3-\nonumber\\&&8z_1^3z_3z_2-z_1^2z_3^2
+2z_2z_3^2z_1+6z_1^2z_2^2z_3+4z_1^3z_2^2z_3+4z_1^3z_2z_3^2-2z_1^2z_2z_3^2)\theta_3^2.
\end{eqnarray}
This operator is rational but really complicated. Of course, this system 
is one example of the extended Picard-Fuchs system for this space, and there may well be a more concise extended Picard-Fuchs system. We note that, as mentioned previously, we arrive at an operator of nearly the same complexity by instead considering a system in which the overall scaling of all the -2 curves are taken to be +1 instead of -1. This is therefore some indication that the -1 factor coming from the topological vertex calculation is not intrinsic to the geometry. 

Finally, we present the B-model Yukawa couplings obtained from the above 
extended Picard-Fuchs system. These Yukawa couplings are indeed transformed 
into the A-model Yukawa couplings (\ref{A}) by the mirror transformation 
(\ref{m}).

{\bf B-model Yukawa coupling of Trivalent Toric Graph}
\begin{eqnarray}
Y_{111} &=& z_{1}(4z_{2}z_{3}-1)^2(16z_{2}^3z_{1}^2z_{3}-4z_{2}^3z_{1}-96z_{2}^2z_{3}^2z_{1}^2+ 
20z_{2}^2z_{3}z_{1}+32z_{1}^2z_{2}^2z_{3}-z_{2}^2-8z_{1}^2z_{2}^2+\nonumber\\&&
16z_{3}^3z_{2}z_{1}^2+20z_{2}z_{3}^2z_{1}+32z_{1}^2z_{2}z_{3}^2-
32z_{1}z_{2}z_{3}-2z_{2}z_{3}+4z_{1}z_{2}+2z_{2}-1-4z_{3}^3z_{1}+\nonumber\\&&
2z_{3}-8z_{1}^2z_{3}^2-z_{3}^2+4z_{1}z_{3})T(z_{1},z_{2},z_{3}),\nonumber\\
&&\nonumber\\
Y_{122} &=& (4z_{2}z_{3}-1)(-1+4z_{1}z_{3})^2(4z_{2}z_{3}^2z_{1}-z_{3}^2-12z_{1}^2z_{3}z_{2}+16z_{1}^2z_{2}^2z_{3}-4z_{2}^2z_{3}z_{1}+z_{1}z_{3}\nonumber\\&&-z_{2}z_{3}+z_{3}-4z_{1}^2z_{2}^2+3z_{1}z_{2}-2z_{1}+2z_{1}^2)z_{2}T(z_{1},z_{2},z_{3}),\nonumber\\&&
\nonumber\\
Y_{123} &=& \frac{1}{2}(4z_{1}z_{2}-1)(4z_{2}z_{3}-1)(-1+4z_{1}z_{3})
(32z_{2}^2z_{3}^2z_{1}^2-16z_{1}^2z_{2}z_{3}^2-16z_{2}^2z_{3}^2z_{1}+\nonumber\\&&
2z_{3}^2z_{1}+2z_{3}^2z_{2}-16z_{1}^2z_{2}^2z_{3}+2z_{1}^2z_{3}+
12z_{1}z_{2}z_{3}-2z_{1}z_{3}+2z_{2}^2z_{3}-2z_{2}z_{3}-z_{3}+1+\nonumber\\&&
2z_{1}^2z_{2}-z_{2}-2z_{1}z_{2}-z_{1}+2z_{2}^2z_{1})T(z_{1},z_{2},z_{3}),
\nonumber\\
&&\nonumber\\
T(z_{1},z_{2},z_{3})&=&1/((4z_{1}z_{2}-1)^2(4z_{2}z_{3}-1)^2(4z_{3}z_{1}-1)^2(4z_{1}z_{2}z_{3}-z_{1}-z_{3}+1-z_{2})).
\end{eqnarray}

\end{document}